\documentclass[aps,pra,twocolumn,showpacs,amsmath,amssymb,preprintnumbers,superscriptaddress,10pt]{revtex4-2}

\usepackage{graphicx}
\graphicspath{{figures/}}
\usepackage{SIunits}
\usepackage{hyphenat}
\usepackage[bookmarks=false]{hyperref}
\usepackage{color}
\usepackage{multirow}
\usepackage[capitalise]{cleveref}
\crefname{section}{Sec.}{Secs.}
\Crefname{section}{Section}{Sections}

\definecolor{pink}{RGB}{255,0,255}

\definecolor{red}{rgb}{1,0,0}

\definecolor{ginger}{RGB}{255,150,0}

\definecolor{blue}{RGB}{63,72,204}

\begin{document}

\title{Security boundaries of an optical power limiter for protecting quantum key distribution systems}

\author{Qingquan Peng}
\affiliation{Institute for Quantum Information \& State Key Laboratory of High Performance Computing, College of Computer Science and Technology, National University of Defense Technology, Changsha 410073, China}
\affiliation{School of Automation, Central South University, Changsha 410083, China}
\affiliation{China Greatwall Research Institute, China Greatwall Technology Group Co., Ltd., Shenzhen 518057, China}

\author{Binwu Gao}
\affiliation{Institute for Quantum Information \& State Key Laboratory of High Performance Computing, College of Computer Science and Technology, National University of Defense Technology, Changsha 410073, China}

\author{Konstantin~Zaitsev}
\affiliation{Russian Quantum Center, Skolkovo, Moscow 121205, Russia}
\affiliation{NTI Center for Quantum Communications, National University of Science and Technology MISiS, Moscow 119049, Russia}
\affiliation{Vigo Quantum Communication Center, University of Vigo, Vigo, Spain}
\affiliation{Escuela de Ingeniería de Telecomunicación, Department of Signal Theory and Communications, University of Vigo, Vigo, Spain}
\affiliation{atlanTTic Research Center, University of Vigo, Vigo, Spain}

\author{Dongyang Wang}
\affiliation{Institute for Quantum Information \& State Key Laboratory of High Performance Computing, College of Computer Science and Technology, National University of Defense Technology, Changsha 410073, China}

\author{Jiangfang Ding}
\affiliation{Institute for Quantum Information \& State Key Laboratory of High Performance Computing, College of Computer Science and Technology, National University of Defense Technology, Changsha 410073, China}

\author{Yingwen Liu}
\affiliation{Institute for Quantum Information \& State Key Laboratory of High Performance Computing, College of Computer Science and Technology, National University of Defense Technology, Changsha 410073, China}

\author{Qin Liao}
\affiliation{College of Computer Science and Electronic Engineering, Hunan University, Changsha, 410082, China}

\author{Ying Guo}
\email{guoying@bupt.edu.cn}
\affiliation{School of Computer Science, Beijing University of Posts and Telecommunications, Beijing 100876, China}
\affiliation{School of Automation, Central South University, Changsha 410083, China}

\author{Anqi Huang}
\email{angelhuang.hn@gmail.com}
\affiliation{Institute for Quantum Information \& State Key Laboratory of High Performance Computing, College of Computer Science and Technology, National University of Defense Technology, Changsha 410073, China}

\author{Junjie Wu}
\email{junjiewu@nudt.edu.cn}
\affiliation{Institute for Quantum Information \& State Key Laboratory of High Performance Computing, College of Computer Science and Technology, National University of Defense Technology, Changsha 410073, China}

\date{\today}

\begin{abstract}
Unauthorized light injection has always been a vital threat to the practical security of a quantum key distribution (QKD) system. An optical power limiter (OPL) based on the thermo-optical defocusing effect has been proposed and implemented, limiting the injected hacking light. As a hardware countermeasure, the performance of the OPL under various light-injection attacks shall be tested to clarify the security boundary before being widely deployed. To investigate the OPL's security boundary in quantum cryptography, we comprehensively test and analyse the behavior of OPL under continuous-wave (c.w.) light-injection attacks and pulse illumination attacks with pulses' repetition rate at $0.5$-$\hertz$, $40$-$\mega\hertz$, and $1$-$\giga\hertz$.
The testing results illuminate the security boundary of the OPL, which allows one to properly employ the OPL in the use cases. The methodology of testing and analysis proposed here is applicable to other power-limitation components in a QKD system.

\end{abstract}

\maketitle

\section{Introduction}
\label{sec:intro}

Quantum key distribution (QKD) based on the laws of quantum mechanics provides a promising path to share a symmetric key between two parties, which has been proven to be informational-theoretically secure~\cite{lo2014,gisin2002,diamanti2016}. 
Thanks to the rapid development of QKD, it has become one of the most mature applications in the field of quantum information, and has gradually become globalized and commercialized~\cite{takesue2007,sibson2017,eriksson2019}. 
However, the imperfection in QKD implementations may disclose loopholes, which are employed by quantum hackers to compromise the practical security of QKD systems~\cite{lutkenhaus2000,makarov2006,fung2007,qi2007,lamas-linares2007,zhao2008,lydersen2010a,lydersen2010b,xu2010,li2011a,wiechers2011,lydersen2011c,lydersen2011b,gerhardt2011,sun2011,jain2011,bugge2014,sajeed2015a,huang2016,makarov2016,huang2018,qian2018,huang2019,huang2020,sun2022,chaiwongkhot2022,huang2022,gao2022}. 
The quantum hacking can be classified to be passive attacks and active attacks. The passive attacks do not change the features of a QKD system. Whereas, a quantum hacker can actively modify the operation mechanism of a QKD system, creating a loophole. These quantum active attacks that usually use laser light as a tool are unpredictable, which threatens the practical security of QKD systems in high-risk level.

The active attacks on the source unit, including laser-seeding attacks~\cite{huang2019,zheng2019}, laser-damage attacks~\cite{Ponosova2022,huang2020}, and Trojan-horse attacks~\cite{lucamarini2015,gisin2006,jain2014}, compromise the security of QKD implementation.
One commonality of these attacks is that the adversary (henceforth called Eve) injects unauthorized light into the QKD's source apparatus to eavesdrop secret key information. 
Taking the laser-seeding attack as an example, Eve injects tailored light into the laser diode of Alice, which results in an increase in the intensity of the emitted optical pulses. 
Under such an attack, the secret key rate of the original protocol may be overestimated, since Alice and Bob are not aware of the existing of the attack~\cite{huang2019}.
This indicates that Eve can actively open security loopholes to successfully obtain information about secret keys in these QKD systems. Considering these attacks in the security proof may eliminate their security threats.

Active attacks on measurement devices, including after-gate attacks~\cite{wiechers2011}, blinding attacks~\cite{wu2020,gao2022,lydersen2010,huang2016} and laser-damage attack~\cite{bugge2014}, are another particularly powerful category of side-channel attacks. 
In these attacks, Eve exploits the imperfection of the single-photon detector, using laser light to control the behavior of the detectors. For example, in blinding attacks, Eve sends relatively strong laser light that makes the detector unable to operate in Geiger mode, no longer sensitive to a single photon~\cite{lydersen2011c}. Applying the blinding attack method, Eve could obtain $100\%$ of the key information, without being noticed by Alice and Bob. Fortunately, innovative QKD protocols, such as measurement-device-independent~(MDI) QKD, are immune to these detection-side-channel attacks~\cite{lo2012}.

While conducting the above-mentioned active quantum attacks, the injected unauthorized light is the essential tool for Eve. This unauthorized bright light modifies the characteristics of the targeted QKD system, breaking some vital security assumptions of QKD protocol~\cite{jain2014,huang2019}. Although we cannot stop Eve injecting hacking light to a QKD system, a module as protection to limit the injected light could be applied, either in the source unit or the measurement unit. This injection-power-limitation module shall present dynamical insertion loss, i.e., the higher the input power the higher the insertion loss. Moreover, this module shall not affect the other properties, except for limiting the injection power. Typically, a module named optical power limiter (OPL), uses nonlinear optical effects to keep the output optical power being stable and below a threshold~\cite{siegman1962}. A preliminary study of an OPL and its application to a QKD system is presented in Ref.~\cite{zhang2021}, which highlights its power-limiting effects. Specifically, when the injected power is between $31~\milli\watt$ to $100~\milli\watt$, its output power can be stabilized around only $1~\milli\watt$.

However, a hardware patching shall be investigate to verified its security performance, to iterate on enhancing security. For each iteration, the security boundary of the patch shall be investigated to show its capability and limitation.
Under this guideline, the security boundary of the OPL that limits eavesdropping light shall be comprehensively studied under various quantum attacks that employ higher power of continuous-wave (c.w.) light and also may use the pulsed light. 
Although some basic testing of OPL has already been conducted in Ref.~\cite{zhang2021}, the security boundary of the OPL shall be comprehensively studied under various quantum attacks that employ higher power of continuous-wave (c.w.) light and also the pulsed light. For this purpose, we tested the OPL in the following hacking scenarios that are not fully investigated in Ref.~\cite{zhang2021}.
Scenario one, Eve illuminates the OPL with high-intensity c.w.\ light, whose optical power is up to $5~\watt$. Scenario two, Eve illuminates the OPL with strong optical pulses. In this case, we have adopted different repetition frequency ($0.5~\hertz$, $40~\mega\hertz$, and $1~\giga\hertz$) of the injected pulses to deeply test the behaviour of the OPL under pulse illumination attack and Trojan-horse attack. The testing results present the security boundary of the OPL, which exposes its limitation of protection for a QKD system. This work provides some reference and guide on properly using the OPL. 
Furthermore, the methodology of testing and analyzing the security boundaries proposed in this work can be applied to other hardware patching. The study aims to gaining a better and comprehensive understanding of the devices in safeguarding QKD systems, supporting the improvement of their security performance. These research findings hold significant value for researchers and practitioners in the fields of information security and quantum communication.

The paper is organized as follows. The \cref{sec:setup} first introduces an experimental model of the OPL and then calibrates the performance of the OPL. The c.w.\ light experiments and pulsed experiments on the OPL are presented in \cref{sec:cw-laser} and \cref{sec:pulse-laser}, respectively. In \cref{sec:discuss}, we discuss the security boundary and make relevant recommendations for using. Finally, the work is concluded in \cref{sec:conclusion}.

\section{Experimental setup}
\label{sec:setup}

\begin{figure*}[htbp]
  \includegraphics[width=1\textwidth]{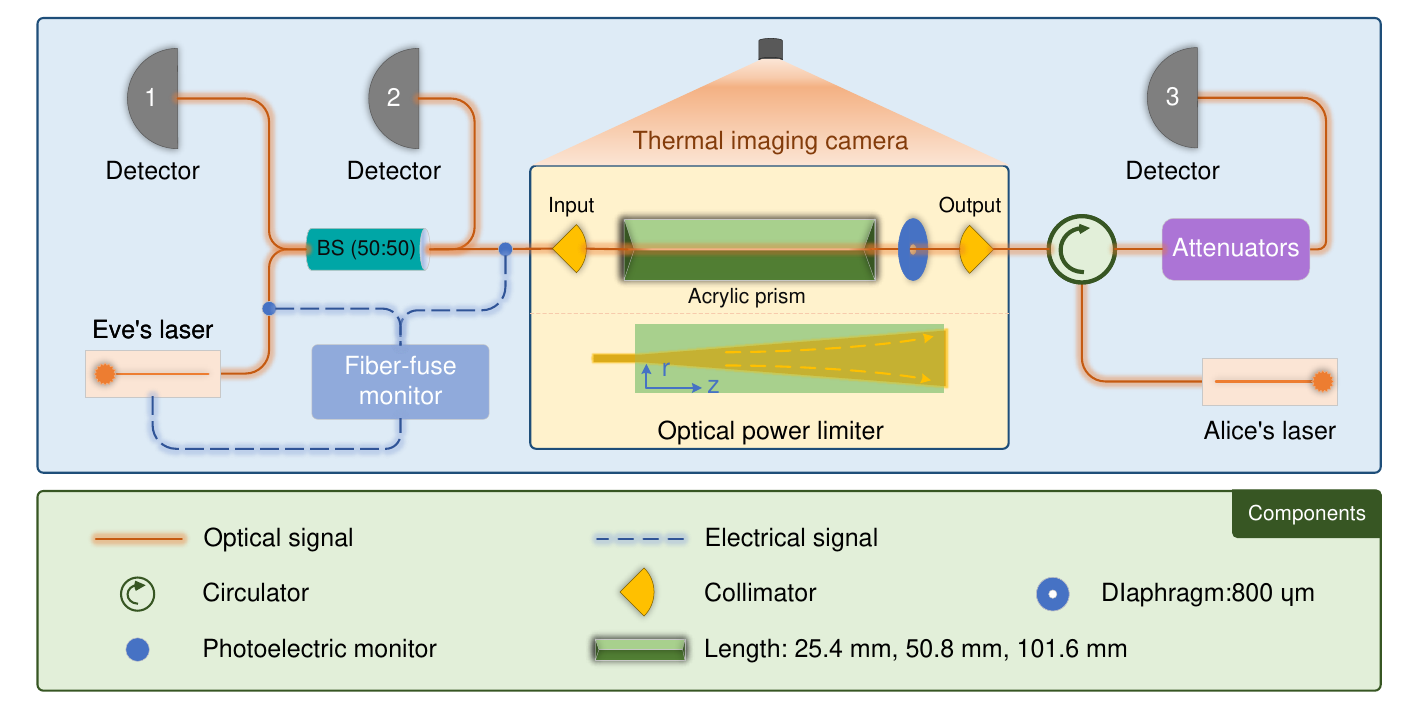}
  \caption{
Experimental scheme of the OPL. The core part of the OPL consists of two collimators, an acrylic prism, and a diaphragm. The internal temperature of the acrylic prism changes when absorbing energy, causing the incident collimated Gaussian beam to diverge due to the thermal-defocusing effects of light. A diaphragm behind acrylic prism controls the collected light power. The sub-inset below the OPL is the light signal divergence diagram inside the acrylic prism. Since acrylic prisms are isotropic, both the optical and thermal responses are axially symmetric along the optical axis. Detector 1 and Detector 2 are optical power detectors. Detector 3 is an optical power meter in the c.w.\ light and $0.5$-$\hertz$ pulsed light experiments. While in the experiments of injection pulses at $40$-$\mega\hertz$ and $1$-$\giga\hertz$ repetition frequencies, an optical-electrical converter is used as Detector 3, and a variable optical attenuator is added. A fiber-fuse monitor protects Eve's laser. A thermal imager records the temperature change of the OPL during the experiment in real time.}
  \label{fig:schematic}
\end{figure*}

\subsection{Scheme of experiment}
The scheme of our experiment is shown in~\cref{fig:schematic}, where the orange solid lines represent the optical signal and the blue dashed lines represent the electrical signal. The OPL is a plug-and-play component, shown as the yellow shaded module in~\cref{fig:schematic}, which consists of two collimators, an acrylic prism, and a diaphragm aperture. In this experiment, three types of acrylic prisms with lengths of $25.4~\milli\meter$, $50.8~\milli\meter$, and $101.6~\milli\meter$ were used, respectively. Diaphragm pore was chosen to be $800~\micro\meter$ in diameter. For the complete testing of power limitations, the experimental setup includes Eve's and Alice's light source. Eve's laser is located at the input side of the OPL to model hacking injected light via a quantum channel. Alice's laser is placed at the other side of the OPL, acting as the laser source used to prepare the weak coherence states in a QKD system.

Eve's laser can produce both pulsed and c.w.\ light. The maximum output power of c.w.\ light reaches $10~\watt$, and the maximum peak power of the pulsed light used in the experiment is $800~\milli\watt$. The laser (c.w.\ or pulse light) produced by Eve is split into two by a $50$:$50$ beam splitter~($50$:$50$ BS). One of the split optical paths is connected to Detector 2, monitoring the output power of Eve's laser in real time. The other half of the split light is injected into the OPL. When passing through the OPL, the laser beam first passes through a circulator and is detected by Detector 3. It is worth noting that Detector 3 refers to an optical power meter used in the following c.w.\ light and $0.5$-$\hertz$ pulsed light experiments but an optical-electrical converter applied in the $40$-$\mega\hertz$ and $1$-$\giga\hertz$ pulse experiments introduced later in the paper. When Detector 3 represents an optical-electrical converter, a variable attenuator is needed to ensure that the power of the input light is in the linear response range of the optical-electrical converter. Alice's laser produces a $3$-$\milli\watt$ c.w.\ light, which passes through the OPL and $50$:$50$ BS and is finally detected by Detector 1.

In order to prevent the setup from damage to the experimental equipment other than the OPL when using the high-power laser, a set of protective measures are also employed. A fiber-fuse monitor containing two fiber-fuse sensors and an automatic shutdown system is applied to protect the setup from fiber fuse that possibly occurs under high-power injection. The fiber-fuse sensors are placed along the fiber jacket, shown as blue dots in~\cref{fig:schematic}. Once fiber fuse is detected, the monitoring circuit automatically shuts down Eve's high-power laser, stopping the fiber fuse and preventing extensive damage to equipment. Fortunately, fiber fuse did not occur during our experiment. In addition, we add a circulator to the output of the OPL to protect Alice's laser from being destroyed by Eve's high-power laser beam.

\subsection{Calibration} 

\begin{figure}[htbp]
\centering
\includegraphics[width=0.5\textwidth]{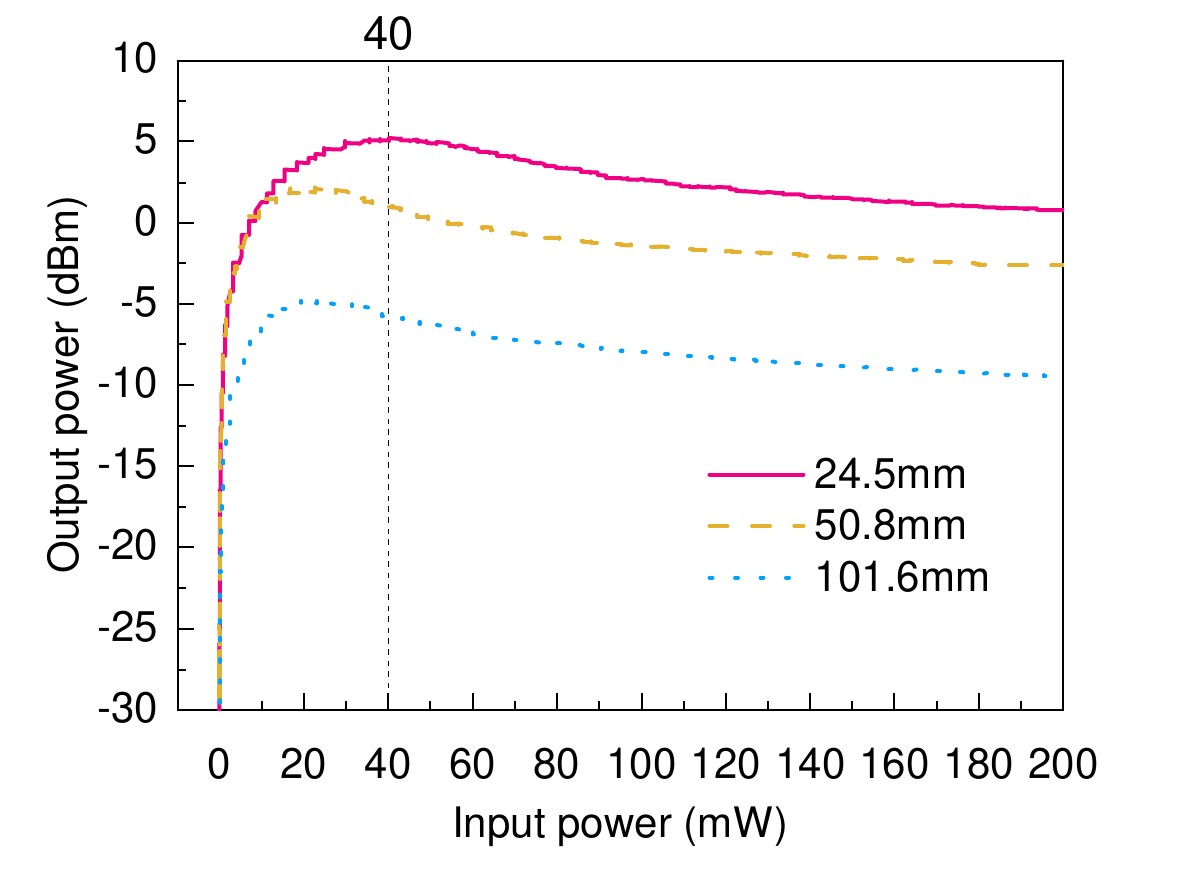}
\caption{
  Calibrated characteristics of power limitation from the forward direction of the optical power limiter. The ``Input power" represents the laser power that Eve injects into the OPL from its input port. The ``Output power" represents the optical power measured at the output port of the OPL.
}
\label{fig:eve-alice}
\end{figure}

\begin{figure}[htbp]
\centering
\includegraphics[width=0.5\textwidth]{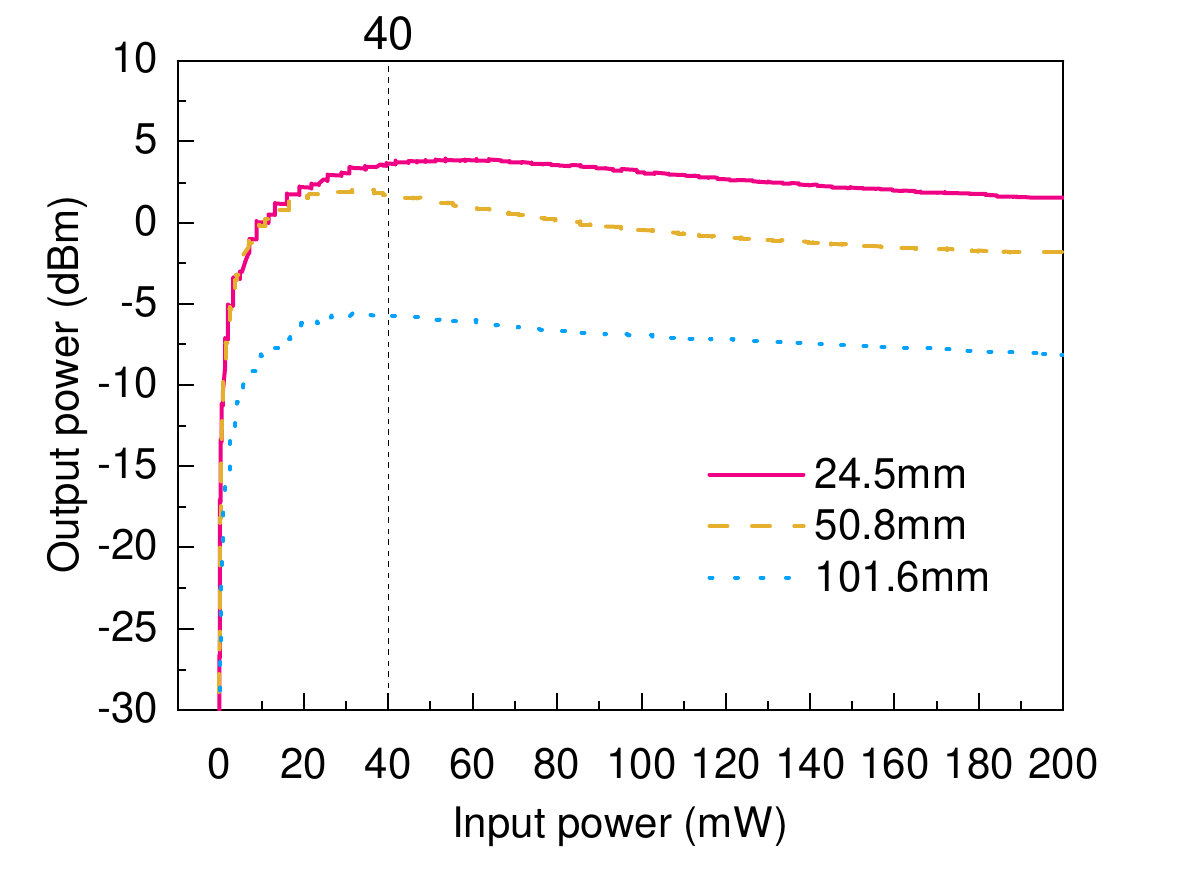}
\caption{
  Calibrated characteristics of power limitation from the backward direction of the optical power limiter. The ``Input power" represents the laser power that Alice injects into the OPL from its output port. The ``Output power" represents the optical power measured at the input port of the OPL.
  }
\label{fig:alice-eve}
\end{figure}

Before the experiment, we used a $1550~\nano\meter$ laser to calibrate the power-limitation feature of the OPL in both forward (input-output) and backward (output-input) directions.
During calibration, we minimize attenuation from the forward direction for each sample. The input power from the forward direction is monitored by Detector 2 and converted to the power at the input port of the OPL by considering the specific splitting ratio of the 50:50 BS. Consequently, the output power is measured by Detector 3 and converted to the power at the output port of the OPL by compensating the insertion loss of the circulator. The calibrated characteristics of power limiter from the forward direction is shown in~\cref{fig:eve-alice}. Similarly, \cref{fig:alice-eve} presents the calibrated property of power limitation from the backward direction. The red solid line, yellow dashed line, and blue dotted line in \cref{fig:eve-alice} and \cref{fig:alice-eve} represent that the lengths of acrylic prisms are $25.4~\milli\meter$, $50.8~\milli\meter$, and $101.6~\milli\meter$. In both~\cref{fig:eve-alice} and \cref{fig:alice-eve}, the functionality of limiting the transmitted optical power is clearly illustrated from both directions, and the output power stays to be stable when in input power is from $40~\milli\watt$ to $200~\milli\watt$ that is the maximal power applied in the calibration. It is also found that the longer the acrylic prism equipped in the OPL provides greater attenuation. For example, in~\cref{fig:eve-alice}, when the input power is $200~\milli\watt$, the output power of the OPL with a $25.4$-$\milli\meter$ acrylic prism is $1.2~\milli\watt$, while that of the OPL with a $50.8$-$\milli\meter$ and $101.6$-$\milli\meter$ acrylic prism is reduced to $551~\micro\watt$ and $114~\micro\watt$, respectively.

\begin{figure}[htbp]
\centering
\includegraphics[width=0.5\textwidth]{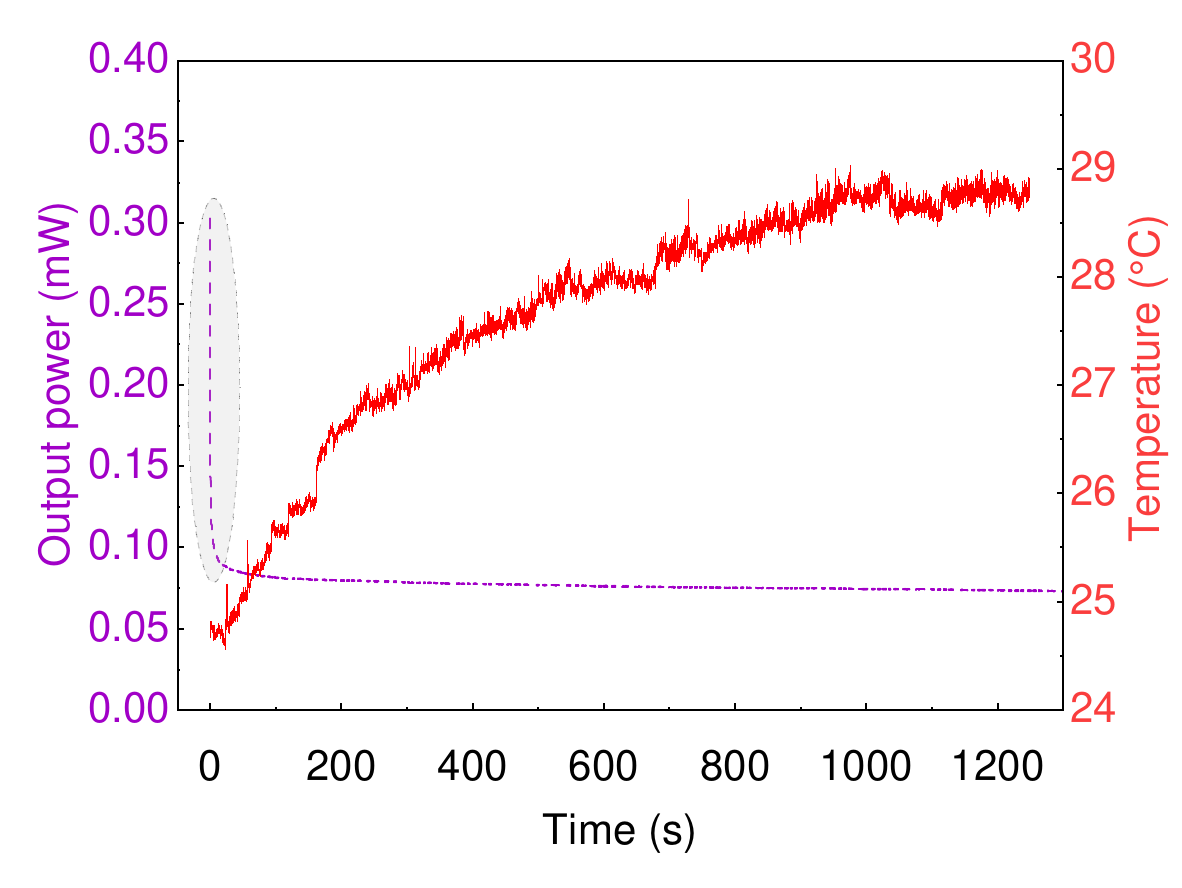}
\caption{
 Variation of output power and temperature during OPL operation in 20 minutes. The input power is $200~\milli\watt$ and the length of the acrylic prism is $101.6~\milli\meter$.
}
\label{fig:longtime}
\end{figure}

The core component of the OPL is the acrylic prism, which effectively controls the output optical power based on the thermo-optical defocusing effect~\cite{smith1977,leite1967,derosa2003}. In order to characterize the stability of power limitation, we sent $200$-$\milli\watt$ c.w.\ light to the OPL from the forward direction for 20 minutes, and obtained the change of output power and temperature over time, as shown in~\cref{fig:longtime}. The results show that the temperature of the acrylic prisms increases rapidly in the first 10 minutes of being irradiated and tends to reach a balance between heat accumulation and heat dissipation in the second 10 minutes. It is worth noting that once we turns on the laser, there is a set of very strong optical power detected by Detector 3, as shown in the grey shaded area in~\cref{fig:longtime}. In other words, the c.w.\ light emitted by Eve is not limited until the OPL activates the thermo-optical defocusing effect. It is shown that the thermo-optical defocusing effect time of acrylic prism is about $200~\milli\second$~\cite{zhang2021}. This time gap may provide some opportunity for Eve to use pulsed light to mitigate the effect of power limitation, which we show in the experiments presented later in this paper.

As typical cases, here we only illustrate the calibration results for three OPLs equipped with different lengths of the acrylic prism. In practice, each OPL used in the following experiment is calibrated according to the above procedure to be precise.

\section{Testing under Eve's c.w.\ laser}
\label{sec:cw-laser}

In principle, Eve is able to hack a QKD system via applying optical power that is limited by the handling capability of the quantum channel~\cite{huang2020}. 
For a fiber based QKD system, Eve’s hacking power shall be under the laser-induced damage threshold (LIDT) of the standard single-mode fiber that acts as the quantum channel. Thus, as the component that Eve’s hacking light first reaches, the OPL’s performance of power limitation shall be experimentally tested under full range of allowed hacking power, instead of only $200~\milli\watt$ as maximum tested in Ref.~~\cite{zhang2021}.
Previous research has shown that $20$-$\meter$ single-mode optical fiber can withstand a $10~\watt$ c.w.\ laser~\cite{huang2020}, which is also the specified maximum power of our laser source used in this study. To fully investigate the characterization of the OPL from both forward and backward directions under Eve's c.w.\ light illumination, we design the testing cycle as follows.

\begin{figure*}[ht]
\centering
\includegraphics[width=1\textwidth]{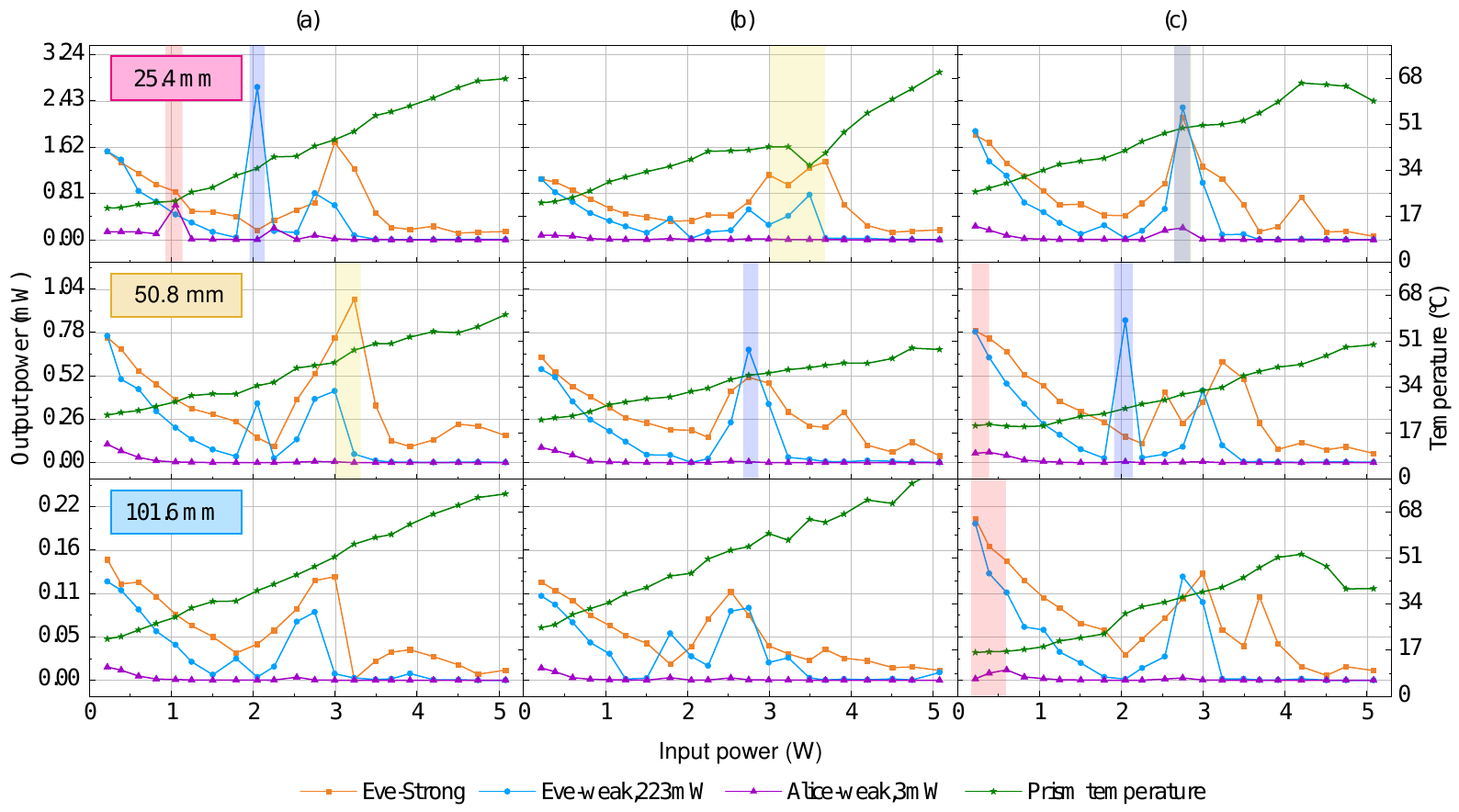}
\caption{The experimental results of OPL shined by Eve's c.w.\ light. ``Input power" refers specifically to the optical power that Eve injects into the OPL laser during the Eve-strong test. ``Output power" corresponds to the output optical power of the OPL during the tested experiment. (a), (b), and (c) are the results of three groups of replicative experiments, and each group contains OPLs with lengths of acrylic prisms are $25.4~\milli\meter$, $50.8~\milli\meter$, and $101.~\milli\meter$.
}
\label{fig:cycle}
\end{figure*}

\begin{enumerate}
  \item [Step 1] Eve's high-power laser testing. Eve injects the c.w.\ light at $1550~\nano\meter$ into the OPL from the forward direction, whose power is $223~\milli\watt + k * 200~\milli\watt$ (here k is cycle number starting from $0$) in the first round. Each test lasts 2 minutes at which point the measurement results are relatively stable. Detector 2 monitors the power injected into the OPL by Eve in real time, and Detector 3 measures the light power passing through the OPL. Meanwhile, Alice's laser is turned off. We name this process as ``Eve-strong test".
  \item [Step 2] Eve's weak-power laser testing. Eve adjusts her laser power to be $223~\milli\watt$ (constant during all cycles) and injects into the OPL again. The injection time lasts $10$-$20$ seconds until the detection value is relatively stable. Detector 3 measures the light power transmitted through the OPL from the forward direction. We name this process as ``Eve-weak test".
  \item [Step 3] Alice's laser testing. Alice generates $3$-$\milli\watt$ c.w.\ light, sending to the OPL from the backward direction. The injection time lasts $10$-$20$ seconds until the detection value is relatively stable. Detector 1 measures the optical power transmitting through the OPL. Meanwhile, Eve's laser is turned off. We name this process as ``Alice-weak test".
  \item [Step 4] Eve increases the laser power she used in Step 1 by $200~\milli\watt$. Then repeat Step 1 to 4 until Eve's injection power reaches $5~\watt$.
\end{enumerate}

According to the above-mentioned test procedure, we have conducted three groups of cyclic experiments. In each group of experiment, the OPLs with three different lengths of acrylic prisms, $25.4~\milli\meter$, $50.8~\milli\meter$, and $101.6~\milli\meter$, are tested. Since acrylic appeared irreversible damage
after each round of experiments, each sample of acrylic prisms can only be tested once. The experimental results are shown in~\cref{fig:cycle}, in which the yellow, blue, and purple lines represent the measured optical power after transmitting through the OPL in the Eve-strong test, Eve-weak test, and Alice-weak test, respectively. During the experiment, the OPL device is placed inside a box surrounded by a black metal plate, but the top of the box is open. A thermal imaging camera looks down on the entire OPL device from the top of the box and records the entire experiment. As shown in~\cref{fig:cycle}, the green stars represent the highest values obtained by the thermal imaging camera in each round of testing. 

Although the OPL under test indeed limits the transmitted power, a more significant variation in power is still shown in certain areas. After being illuminated by Eve's light up to $1~\watt$, the transmittance of the OPL from the backward direction increases slightly, as shown by the purple lines in the red shaded areas in~\cref{fig:cycle}. We take the OPL with the acrylic prism length of $25.4~\milli\meter$ in group (a) as an example. When Alice's injection power keeps being stable at $3~\milli\watt$, the transmitted power from the backward direction increases from $0.13~\milli\watt$ to $0.61~\milli\watt$ after shining by Eve's $1$-$\watt$ light. The similar phenomenon also appears in the OPLs tested in group (c). That is, the output power increases from $0.057~\milli\watt$ to $0.063~\milli\watt$ in the sample with $50.8$-$\milli\meter$-long acrylic prism, after being shined by $0.388~\watt$ light from the forward direction. For the OPL with $101.6$-$\milli\meter$-long acrylic prism, the output power increased from $0.002~\milli\watt$ to $0.012~\milli\watt$ under $0.6$-$\watt$ Eve's injection power.

The transmittance from forward direction under Eve-weak testing becomes higher than its original value after the OPL is shined by Eve's light with power between $2$ to $3~\watt$, as shown by the blue lines in the shaded areas in~\cref{fig:cycle}. Taking group (a) as an example, in the sample with the acrylic prism length of $25.4~\milli\meter$, the optical power injected into the OPL by Eve's laser is stable at $223~\milli\watt$, but the output power is increased from $1.5464~\milli\watt$ to $2.672~\milli\watt$. The same phenomenon also appeared in groups (b) and (c). For samples with acrylic prism length of $50.8~\milli\meter$ in groups (b) and (c), the output power increased from $0.56~\milli\watt$ to $0.677~\milli\watt$ and $0.784~\milli\watt$ to $0.853~\milli\watt$, respectively. For sample with acrylic prism length of $25.4~\milli\meter$ in group (c), the output power increased from $1.9~\milli\watt$ to $2.315~\milli\watt$.

As the injected optical power increases from $2~\watt$ to $4~\watt$, the transmitted power under Eve-strong testing also increases, as shown by the orange line in the yellow shaded and gray shaded area in \cref{fig:cycle}. In group (a), the output power of the sample with the acrylic prism length of $50.8~\milli\meter$ increases from $0.76~\milli\watt$ to $0.98~\milli\watt$. Similar, for the samples with acrylic prism length of $25.4~\milli\meter$ in groups (b) and (c), the output power increases from $1.06~\milli\watt$ to $1.366~\milli\watt$ and $1.83~\milli\watt$ to $2.315~\milli\watt$, respectively. In general, the longer the acrylic prism, the lower the probability of increased output power. For example, when the acrylic prism is $101.6~\milli\meter$ in group (a) and (b), no transmitted power of the OPL is raised. Once the injected power is beyond $4~\watt$, the transmittance drops due to the physical damage.

\begin{figure}[ht]
\centering
\includegraphics[width=0.5\textwidth]{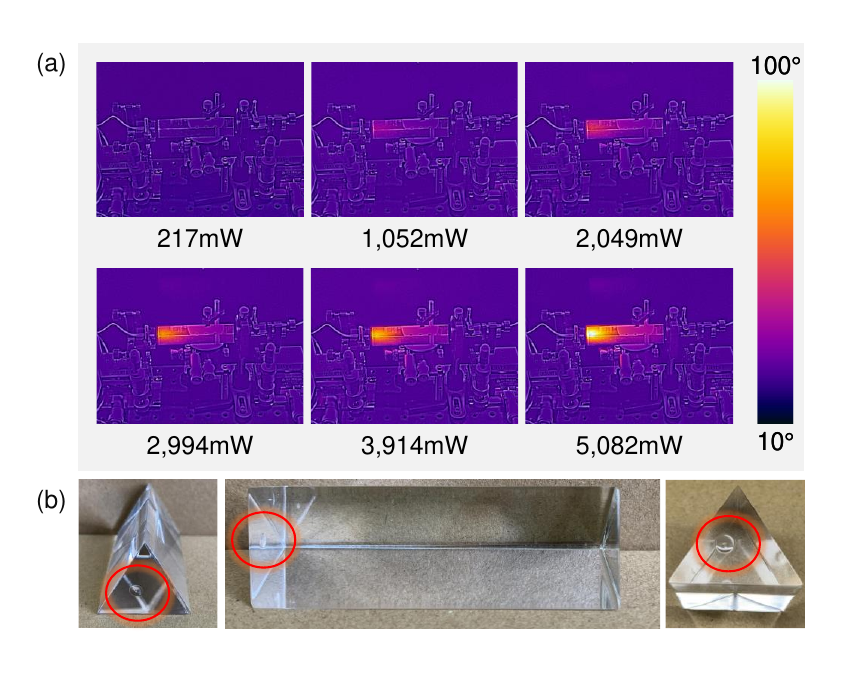}
\caption{Analysis of the OPL response to high-power laser exposure. (a) Thermal image of an acrylic prism illuminated by a high-power c.w.\ laser. (b) Photographs of the acrylic prism after experiment. The length of the acrylic prism is $101.6~\milli\meter$. The red marks indicate the positions of raised bubbles after the acrylic prism was damaged. }
\label{fig:thermal}
\end{figure}

Each acrylic prism sample suffers irreversible physical damage after completing the cycling test. The damaged OPL has a very high optical attenuation, transmitting only a few tens of $\nano\watt$ optical power under $5$-$\watt$ injection light. In order to further understand this damage, we used a thermal imager to continuously observe the temperature of the OPL, which is shown in~\cref{fig:thermal}(a). The power value marked under each thermal image represents the optical power injected into the OPL under Eve-strong testing. It can be seen that the heat accumulates in the acrylic prism as the injected optical power continues to increase. Thus, the overall temperature of the OPL begins to rise due to the thermal radiation of the acrylic prism. The highest temperature point is right behind the contact surface between the acrylic prism and the collimator at the input port. \Cref{fig:thermal}(b) shows the photographs of acrylic prism after being tested. A small bubble appeared on the surface where Eve's laser beam first reaches, which shows that the high-power laser causes permanent damage on the OPL.

\section{Testing under Eve's pulsed laser}
\label{sec:pulse-laser}

In quantum attacks, Eve is not only allowed to apply c.w.\ light to QKD system~\cite{lydersen2010, bugge2014, huang2016, huang2020}, but also can use optical laser pulses to exploit loopholes~\cite{gisin2006, wu2020, gao2022, lucamarini2015}. It shows that the threat caused by pulsed light is as much as that caused by c.w.\ light, which is because some attacks relies on the instantaneous energy instead of average energy. Therefore, we extend the investigation of the OPL’s property of power limitation to optical pulses with repetition rate of $0.5~\hertz$, $40~\mega\hertz$, and $1~\giga\hertz$ in this section.

\subsection{Eve's optical pulses with $0.5$-$\hertz$ repetition rate} 
\begin{figure}[ht]
\centering
\includegraphics[width=0.5\textwidth]{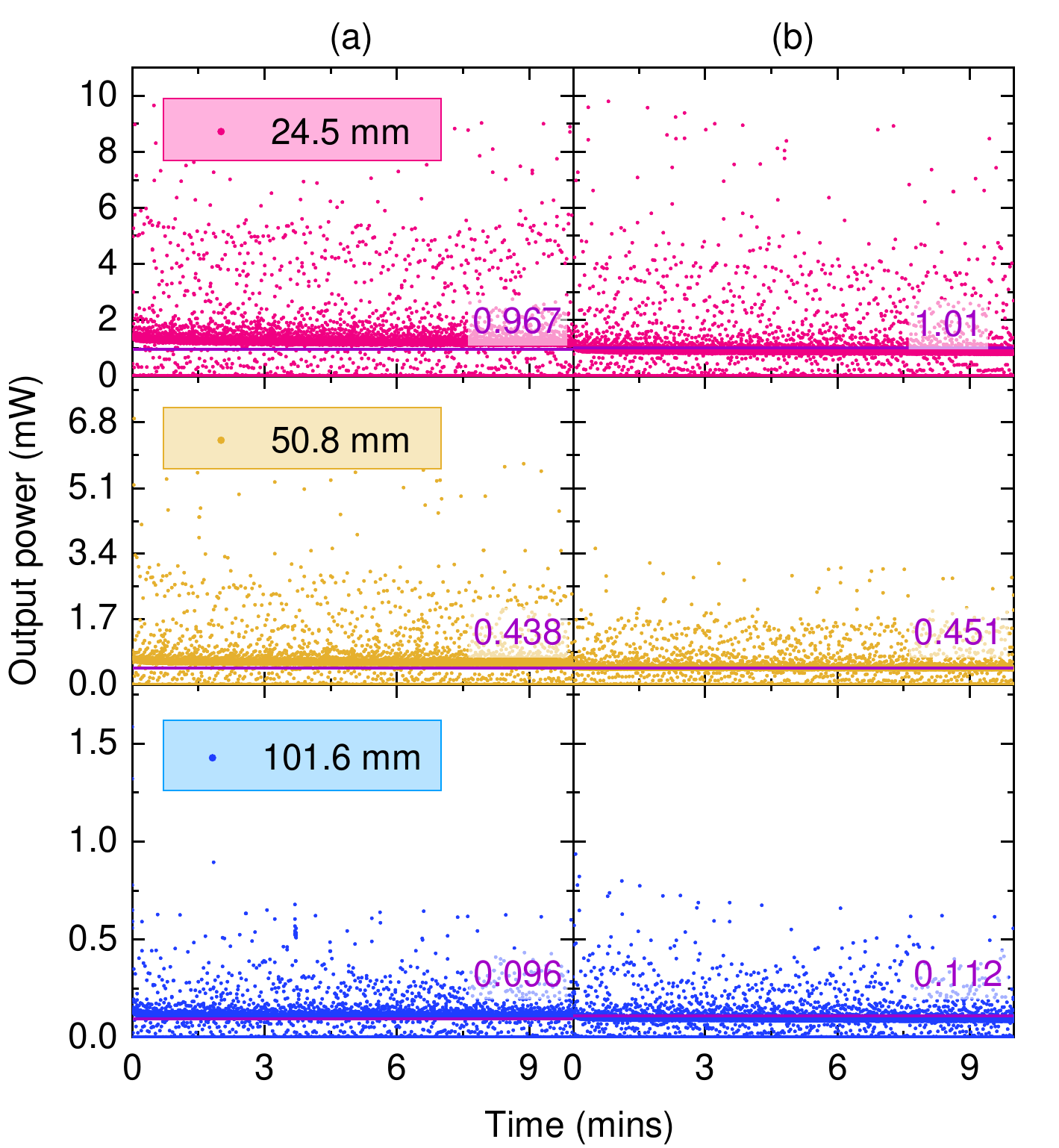}
\caption{The scatter plots of the OPL's output power under Eve's injection pulses with repetition frequency of $0.5$-$\hertz$. (a) The peak power of Eve's optical pulses is $200~\milli\watt$. (b) The peak power of Eve's optical pulses is $400~\milli\watt$.
}
\label{fig:pulselongtime}
\end{figure}
\begin{figure*}[ht]
\centering
\includegraphics[width=1\textwidth]{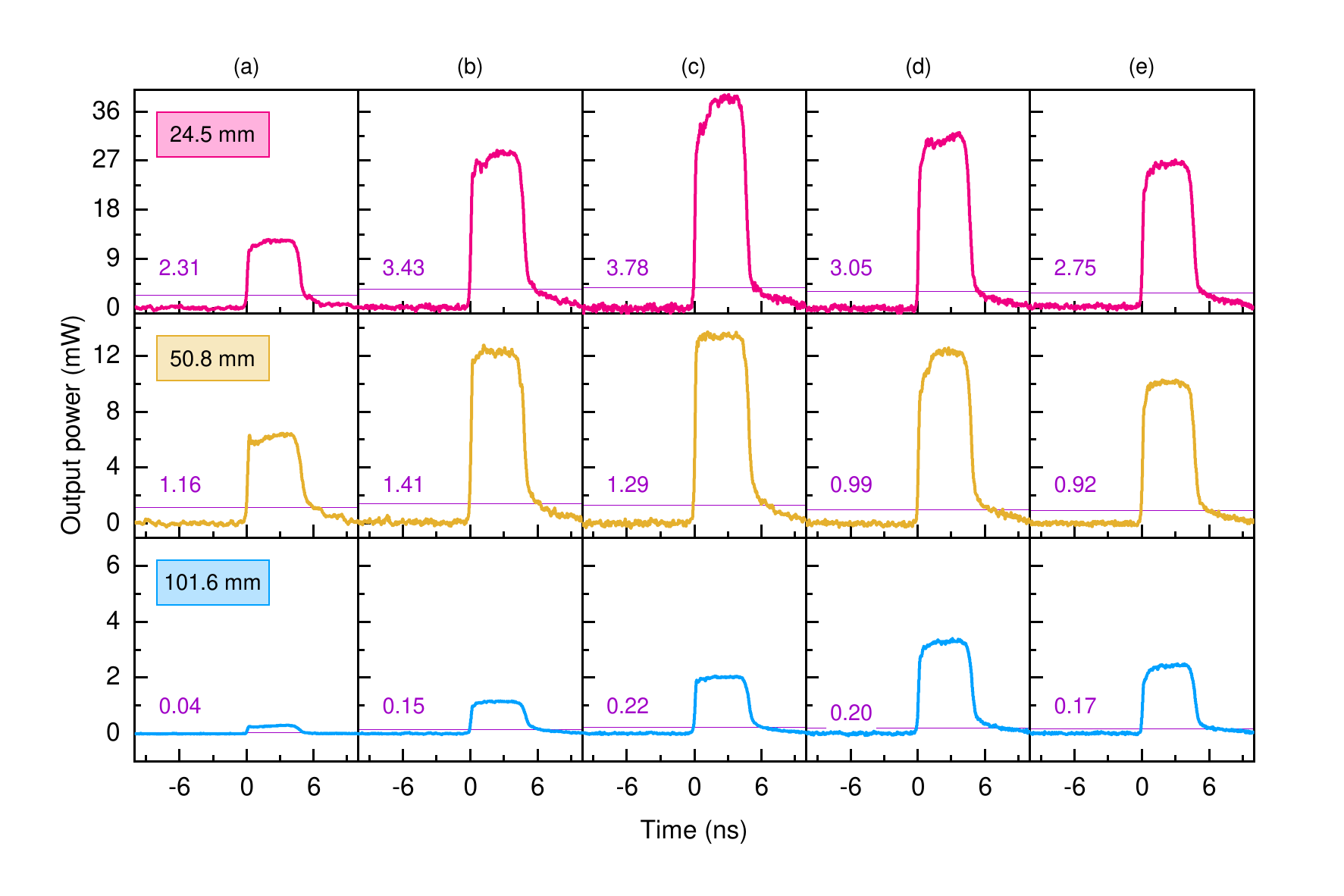}
\caption{Output waveforms of the OPL under $40$-$\mega\hertz$ pulsed light. The average optical power injected by Eve into the OPL is (a) $10~\milli\watt$, (b) $20~\milli\watt$, (c) $30~\milli\watt$, (d) $60~\milli\watt$, and (e) $80~\milli\watt$.}
\label{fig:pulse_40mhz}
\end{figure*}

During the calibration, \cref{fig:longtime} shows that the transmitted power reaches to a peak right after the optical power sending to the OPL and then drops to a steady level. This indicates that the capability of power limitation may vary over time and relate to the illumination duration. 
In order to study the power limitation response under difference illumination duration, we first apply optical pulses working at $0.5$-$\hertz$ repetition rate and wavelength of $1550\nano\meter$ to the OPL as initial trial. The duty cycle of the light pulse light is 50\% and the peak power is $200~\milli\watt$ and $400~\milli\watt$ respectively. The output power is recorded every $0.1~\second$ by Detector~3 over 10 minutes to be a set of scatter plots. 

\Cref{fig:pulselongtime} shows the power limitation response of the OPL illuminated by Eve's optical pulses working at the repetition frequency of $0.5$-$\hertz$. The red, orange, and blue scatter plots in the figure represent the output power of the OPL with $25.4$-$\milli\meter$-long, $50.8$-$\milli\meter$-long, and $101.6$-$\milli\meter$-long acrylic prisms, respectively. The phenomenon that the longer the acrylic prism leads to the lower the output power is also shown in this testing. In addition, the purple horizontal line in each prism diagram indicates the optical power detected by Detector 3 if Eve apples c.w.\ light with power as the same as optical pulses' peak power. It can be seen from~\cref{fig:pulselongtime} that most measured values of output power under $0.5$-$\hertz$ pulsed laser exceed the purple line. Crucially, this suggests that Eve could exploit the finite response time of the OPL to inject more light into a QKD system. It is also notable that the longer acrylic prism limits the peak power to be lower and closer to the purple line.

\subsection{Eve's optical pulses with $40$-$\mega\hertz$ repetition rate} 
\label{subsec:40mhz}
\begin{figure*}[ht]
\centering
\includegraphics[width=1\textwidth]{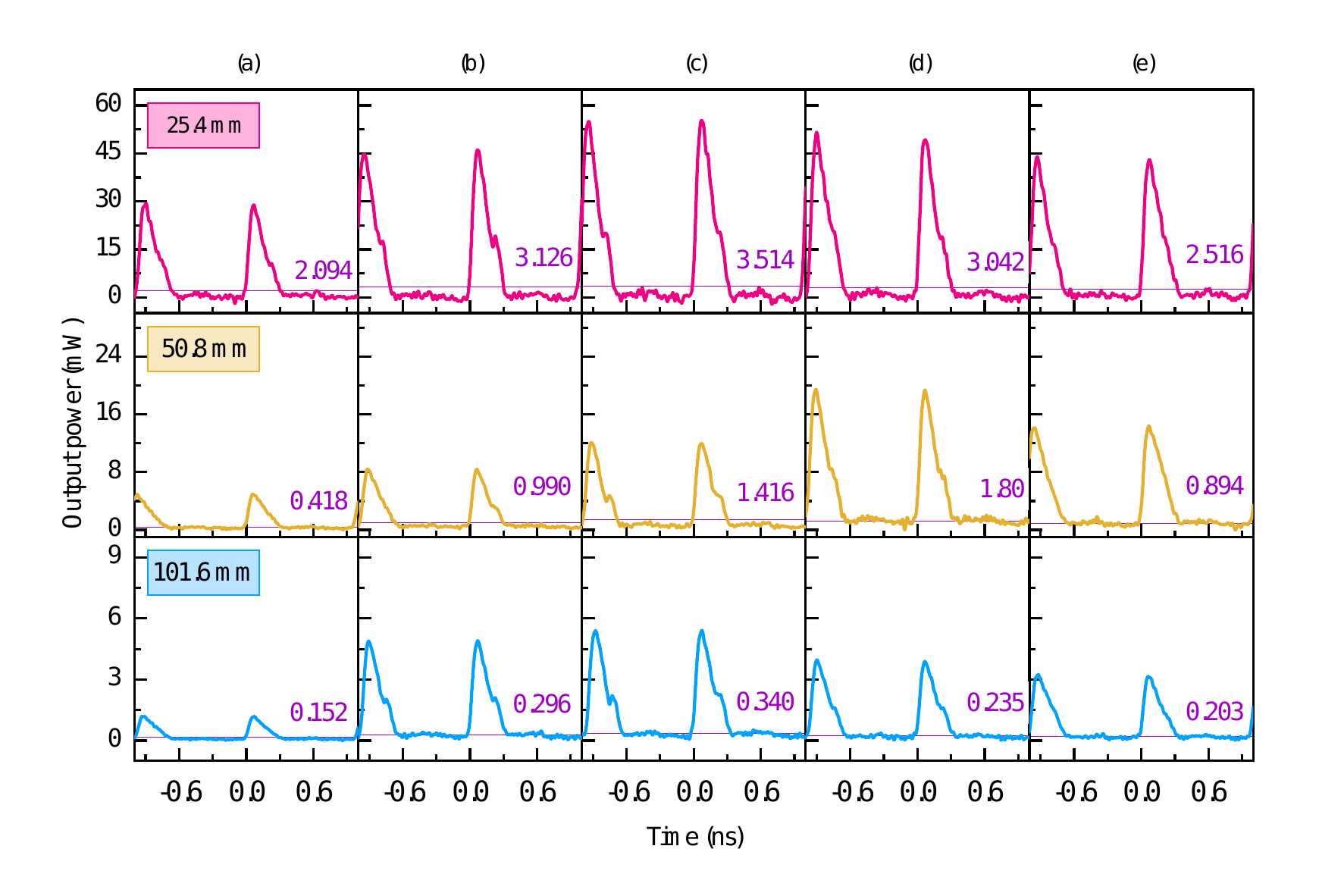}
\caption{Output waveforms of the OPL under $1$-$\giga\hertz$ pulsed light. The average optical power injected by Eve into the OPL is (a) $10~\milli\watt$, (b) $20~\milli\watt$, (c) $30~\milli\watt$, (d) $60~\milli\watt$, and (e) $80~\milli\watt$.}
\label{fig:pulse_1ghz}
\end{figure*}

In a quantum attack, Eve may apply optical pulses with the repetition rate as the same as that of a QKD system, such as the pulse illumination attack presented in Ref.~\onlinecite{wu2020,gao2022}. Restricting the power of Eve's illumination light to the avalanche photodiode~(APD) is one of the most direct and effective means of preventing such attacks. Therefore, it is necessary to investigate whether the OPL can protect APD from these attacks. In this subsection, we tested the transmittance of the OPL under light injection by Eve who employs optical pulses with the same repetition rate as the one used in Ref.~\onlinecite{wu2020}.

Eve's laser emits optical pulses at $1550~\nano\meter$ and its repetition frequency is  $40$-$\mega\hertz$, as the same as that of the APD tested in Ref.~\onlinecite{wu2020}. In our experiments, the width of each optical pulse is set as $4~\nano\second$. Detector 2 monitors the average power of Eve's optical pulses. The attenuator dynamically adjusted its attenuation according to Eve's injection power, ensuring Detector 3 response linearly. We test the performance of the OPL when average power of Eve's optical pulses is $10~\milli\watt$, $20~\milli\watt$, $30~\milli\watt$, $60~\milli\watt$, and $80~\milli\watt$. Additionally, we test the OPL with different lengths of acrylic prisms under each average power of Eve's light.

The testing results are shown in~\cref{fig:pulse_40mhz}. It is clear that while Eve increases the average power of the optical pulses, the output power of the OPL first increases and then decreases, no matter what the length of acrylic prism is used. During the process of Eve increasing the pulse power, the maximal output power of the OPL can be observed when the average power of the optical pulses is $30$ or $60~\milli\watt$. For the OPL with $25.4$-$\milli\meter$-long acrylic prism, the maximal output power of the OPL measured in the experiment is $56.59~\milli\watt$. For the OPL with $50.8$-$\milli\meter$-long/$101.6$-$\milli\meter$-long acrylic prism, the maximal output power of the OPL measured in the experiment is $38.83~\milli\watt$/$13.69~\milli\watt$. Thus, the longer acrylic prism limits the output power to be the lower value. 

In addition, the pulsed light can be easily passed through the OPL compared to the c.w.\ light. The purple line in each subfigure in~\cref{fig:pulse_40mhz} represents the output power of the OPL under c.w.\ light injection whose power is as the same as the average power of the optical pulses. The peak power of the transmitted pulses in each test demonstrates much higher value than the purple line. For example, in \cref{fig:pulse_40mhz}(c), for the OPL equipped with $25.4$-$\milli\meter$-long acrylic prism, Eve's optical pulses result in the transmitted power of $38.83~\milli\watt$, while only $3.78$-$\milli\watt$ optical power transmitted through the OPL under the c.w.\ laser test. In the test, the peak power of the optical pulses is $5.5656$-$16.969$ times higher than that of the c.w.\ light under the same input optical power in average.

\subsection{Eve's optical pulses with $1$-$\giga\hertz$ repetition rate} 
\label{subsec:1GHz}

In this subsection, we investigate the OPL's response of power limitation under high-speed pulse illumination, which might be used in Eve's attack on a high-speed QKD system~\cite{gao2022}. Usually, a high-speed QKD system would have a $\giga\hertz$-level repetition frequency. 
Therefore, in this test, we prepared an optical pulse with $1550~\nano\meter$, $1$-$\giga\hertz$ repetition frequency, and about $200$-$\pico\second$ full width at half maximum as a typical case.

The specific experimental procedure is similar to the \cref{subsec:40mhz}. Each set of experiment consisted of OPLs equipped with acrylic prisms in three different lengths. The average optical power of the pulse injected by Eve into the OPL was gradually increased from $10~\milli\watt$ to $80~\milli\watt$, as the same as that of the last test in~\cref{subsec:40mhz}. The output power is first measured by Detector 3 and then is converted to the power at the output port of the OPL. The testing results are presented in~\cref{fig:pulse_1ghz}. In each subfigure, the purple solid line represents the output power of the OPL if the c.w.\ light injection is performed with optical power as the same as the average power of the optical pulses. From this test, the following phenomena have been disclosed.

Similar to $40$-$\mega\hertz$ optical pulses test, the maximum output power in this test can be obtained when the average optical power injected by Eve into the OPL is $30$ or $60~\milli\watt$. In the OPL equipped with an acrylic prism length of $25.4~\milli\meter$/$101.6~\milli\meter$, the maximum output power is $55.31~\milli\watt$/$5.41~\milli\watt$ when the average input power is $30~\milli\watt$. The maximum output power of the OPL with $50.8$-$\milli\meter$-long acrylic prism is $19.33~\milli\watt$ under input optical pulses with average power of $60~\milli\watt$.

The output power of optical pulses is higher than that of the c.w.\ light injection, when the average input power of the optical pulses is as the same as that of the c.w.\ light. For example, when Eve applies optical pulses with an average optical power of $10~\milli\watt$ to the OPL with a $25.4$-$\milli\meter$-long acrylic prism, its output peak optical power is $28.87~\milli\watt$. Whereas, the output power is only $2.094~\milli\watt$ under $10~\milli\watt$ c.w.\ input light. All the cases that we tested follow this regularity.
  
When the average optical power is constant, the peak output power of the OPL increases with the pulse light frequency. For example, compare the same situation in \cref{fig:pulse_40mhz}~(c) and \cref{fig:pulse_1ghz}~(c), when the length of the acrylic prism assembled by OPL is $25.4~\milli\meter$. We find that the maximum output power corresponding to the pulsed light at $40$-$\mega\hertz$ is $39.6046~\milli\watt$, while the maximum output power corresponding to the $1$-$\giga\hertz$ light is $56.869~\milli\watt$. In this test, the peak power of the optical pulses is $5.627$-$16.969$ times higher than that of the c.w.\ light assuming the averaged input power is the same.

In the pulsed-laser experiments demonstrated above, it is shown that the peak output power of the pulsed light transmitted through the OPL is higher than that of c.w.\ light, assuming the input power in average is equal. The reason for this phenomenon is that the OPL limits the output optical power based on the thermo-optical defocusing effect. In the case of the pulsed light, the OPL only receives a short light radiation followed by a long idle time as a repetition cycle, which is difficult to accumulate the heat to produce the thermo-optical defocusing effect. Thus, more light transmitted through the OPL under the pulsed injection.

\section{Discussion about security boundary}
\label{sec:discuss}

In the tests shown above, the OPL indeed demonstrates its power-limitation effect. Besides of that, the OPL also shows variation in response under c.w.\ high-power light and pulsed light. 
From the results and principles of OPL, it can be seen that the length and temperature changes of acrylic prism will affect its power limiting performance. Therefore, the security boundary of the OPL is studied under potential risk of quantum attacks, considering parameters of the acrylic prism for necessity.

\subsection{Security boundary for c.w.\ light injection}

In the c.w.\ light experiments, generally, the OPL keeps the transmitted power to be the same order of optical power no matter how much input light is. Nevertheless, the OPL still may expose some security threats that shall be noticed. This is because that the significant increase in optical power allows us to observe the performance fluctuation and degradation of the OPL under high-intensity light, which also was not explored in the previous work~\cite{zhang2021}, presenting the security boundary of the OPL.

Firstly, for the source unit of a QKD system, the amount of Alice's transmitted light slightly increases as shown in our test, which may let the mean photon number be higher than the set value. Usually, in Alice, the insertion loss of the OPL is counted as a part of attenuation that attenuates the weak coherence laser to be the single-photon level. In normal case, the mean photon numbers of Alice's output follow the set values required by a QKD protocol. However, if Eve injects about $1~\watt$ c.w.\ light, there is a chance to increase the mean photon number of Alice's pulses as shown by the red area in~\cref{fig:cycle}. This unnoticed increase of the optical power sent by Alice results in the incorrect estimation of the key rate, which compromises the security of a QKD system, either running a prepare-and-measure QKD protocol or a measurement-device-independent QKD protocol~\cite{huang2019}.

\begin{figure}[ht]
\centering
\includegraphics[width=0.5\textwidth]{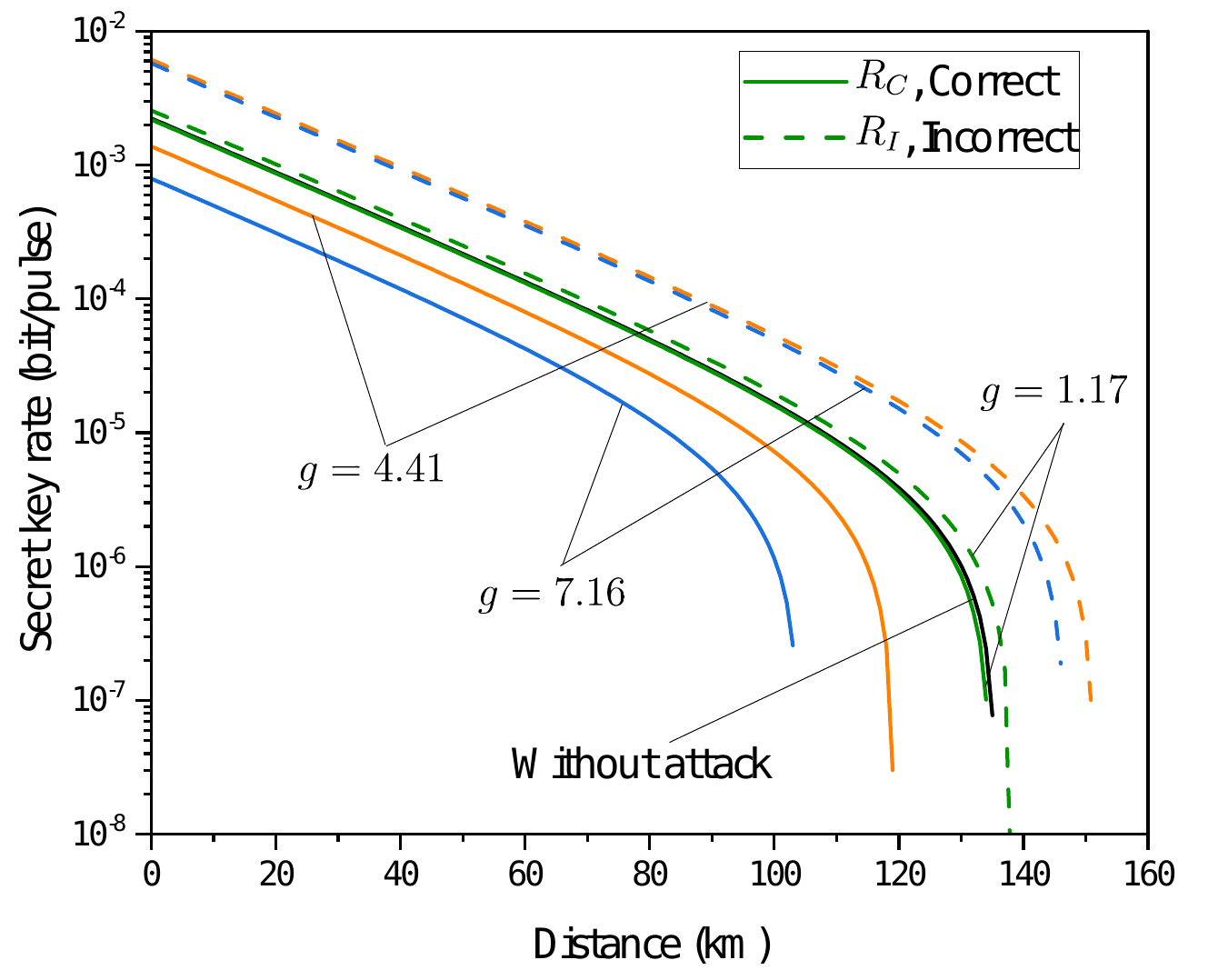}
\caption{Secret key rate as a function of the distance for the standard decoy-state BB84 protocol at different mean photon numbers of Alice's pulses. The average number of photons in the signal state and the decoy state are set to be $0.1$ and $0.05$, respectively. Other parameters in the simulation are as follows. Dark count probability per gate is $2 \times 10^{-6}$; error correction efficiency is $1$; the probability that a photon hit the erroneous detector is $0.01$; Bob's internal optics loss $0.4$; detection efficiency of Bob's single-photon detector is $10\%$.}
\label{fig:bb84}
\end{figure}

Specifically, as shown in \cref{fig:bb84}, the dashed and solid lines represent the incorrect and correct secret key rates for a decoy-state BB84 QKD system with increased in Alice's mean photon number, denoting as $R_I$ and $R_C$, respectively. The green, orange, and blue curves indicate that mean photon numbers of Alice's pulses are increased to $g=1.17$ times, $g=4.41$ times, and $g=7.16$ times of the set values, according to the testing results shown in~\cref{fig:cycle}. The black solid line represents the secret key rate without attack. It is apparent that the secret key rate estimated by Alice and Bob $R_I$ given by the dashed lines is significantly higher than the accurate key rate $R_C$. More precisely, once the attenuation value of the OPL decreases due to the attack, Alice and Bob are not aware of the increase in the transmitted mean photon number and wrongly estimate the secret key according to the security proof introduced in Refs.~\cite{gottesman2004,lo2005}. Remarkably, the dashed lines are higher than the black solid line, which illustrates that the incorrectly estimated secret key rate cannot be guaranteed as a secure one.

\begin{figure}[ht]
\centering
\includegraphics[width=0.5\textwidth]{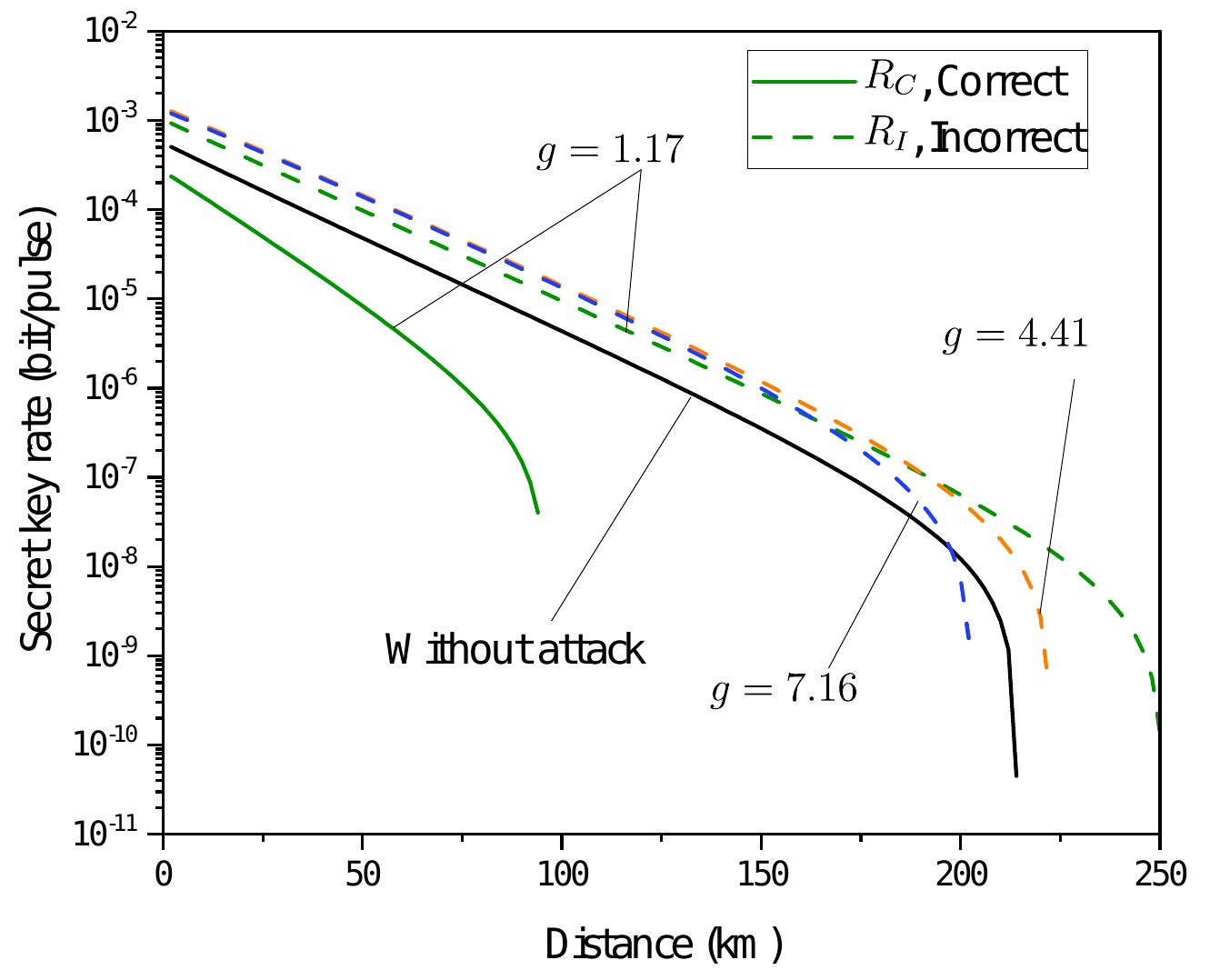}
\caption{Secret key rate as a function of the distance for symmetric MDI QKD protocol at different mean photon numbers of Alice's pulses. The parameters used in the simulations are taken from Ref.~\onlinecite{ma2012}. Alice and Bob set light intensity of the vacuum state, decoy state, and signal state is $0$, $0.03$, and $0.5$, respectively.}
\label{fig:mdi}
\end{figure}

We also consider the case of MDI QKD with weak coherent pulses~\cite{lo2012}. Similar to the previous example, we assume that the attenuation value decreases after the OPL is being attacked, so that the output intensity of both Alice and Bob increases. The resulting secret key rates are shown in \cref{fig:mdi}. From the overall observation, we find that the results are analogous to those illustrated in \cref{fig:bb84}. In particular, in the presence of an attack, the $R_I$ estimated by Alice and Bob are much higher than the correct one $R_C$, provided that Alice and Bob are not aware of the increase in intensity under the attack. It is worth noting that, when the light intensity is  increased by $4.41$ and $7.16$ times, there is no secret key can be generated.

Besides, we can also see from \cref{fig:cycle} that at very beginning of experiment with slow temperature rise there is small decrease in optical power of Alice's pulses. Thus, the lowest temperature corresponds to highest optical power of Alice's pulses. When considering the security of a QKD system with the OPL, temperature manipulation may cause a loophole. For a room placed a QKD system with temperature $20~\celsius$, Eve may try to manipulate conditioner in a lab to decrease its temperature, say to $15~\celsius$. In this case, the decreased temperature may let the optical power of Alice's pulses be higher, which again leads to unnoticed increase of the mean photon number sent by Alice. It is notable whether OPL has the above-mentioned vulnerabilities still needs further experimental verification.

To counter this potential threat, one possible solution is to characterize the minimal insertion loss of the OPL and always take this minimal value into account to calculate the required attenuation for the set mean photon number in Alice. Of course, this countermeasure provides a conservative estimation of the secret key rate, which is lower than the optimal one. Another more conservative but security countermeasure is to assume the OPL in Alice to be untrusted. Thus, the attenuation needed to prepare quantum states with set mean photon number shall be accounted without the OPL. In this case, no matter how the insertion loss of OPL changes, the security assumption about the mean photon number at the set single-photon level would not be affected. Whereas, the payment is the lower secret key rate and the shorter communication distance for a QKD system.

Secondly, for the detection unit of a QKD system, the OPL still can pass through $\milli\watt$-level light under several-hundreds-$\milli\watt$ or $2$-$4$~$\watt$ injected light, which usually is sufficient to conduct a detector blinding attack on an APD. Previous studies have been shown that $\nano\watt$ to $\milli\watt$ c.w.\ light is able to blinding the APD, no matter it is passively quenched module~\cite{gerhardt2011,makarov2009}, actively quenched module~\cite{gras2020}, or the gated module~\cite{chistiakov2019,huang2016,lydersen2010}. Thus, the OPL cannot prevent a QKD system from these blinding attacks. An exception, as mentioned in Ref.~\onlinecite{zhang2021}, is an APD with shorten impedance of the bias voltage supply, which requires about $10.34~\milli\watt$ to blind the APD detector~\cite{gao2022}. This is because the blinding light is no longer drop the bias voltage of this low-impedance biased APD. Instead, the blinding light is used to accumulate the heat in the APD to increase its temperature, increasing the temperature-dependent breakdown voltage~\cite{lydersen2010b}. Apparently, this thermal effect requires more injection light.

However, the OPL is likely to prevent the source unit in a QKD system from the laser-seeding attacks~\cite{sun2015,huang2019} and the laser-damage attacks~\cite{makarov2016,huang2020, Ponosova2022}. Since the injection power is at mW level after the OPL, the laser-seeding attack would not work if Alice applies more than $40$-$\deci\bel$ attenuation to make less than $100$-$\nano\watt$ light injected into the laser diode. Regard to the laser-damage attacks, it is shown in our test that the high-power optical light can be attenuated to several milliwatts, which usually is not enough to damage the optical components in Alice. The high-power optical light even may destroy the OPL to block the transmission, protecting other components behind it from being hacked. Moreover, we find that the c.w.\ high-power light causes a significant increase in the temperature of the OPL. Thus, the OPL combined with a thermal sensor may be able to sense the increased temperature and trigger further protective action once the temperature is beyond a threshold.

\subsection{Security boundary for pulsed light injection}

In the pulsed light experiments, the optical pulses transmitted through the OPL. After that, the peak power of the optical pulses varies from  $0.370$-$56.599~\milli\watt$, which is $5.627$-$16.943$ times than that of the c.w.\ light. The peak power of the transmitted pulses indeed reduces to be much lower than the original pulses. However, the remaining optical pulses still may help Eve to conduct pulse illumination attacks on single-photon avalanche diodes (SPAD) and Trojan-horse attacks on the modulators. In this subsection, we focus on the security boundary in the scenarios of pulse-based attacks equipped with the OPL, providing a more comprehensive picture about the security performance of the OPL.

\begin{figure}[ht]
\centering
\includegraphics[width=0.5\textwidth]{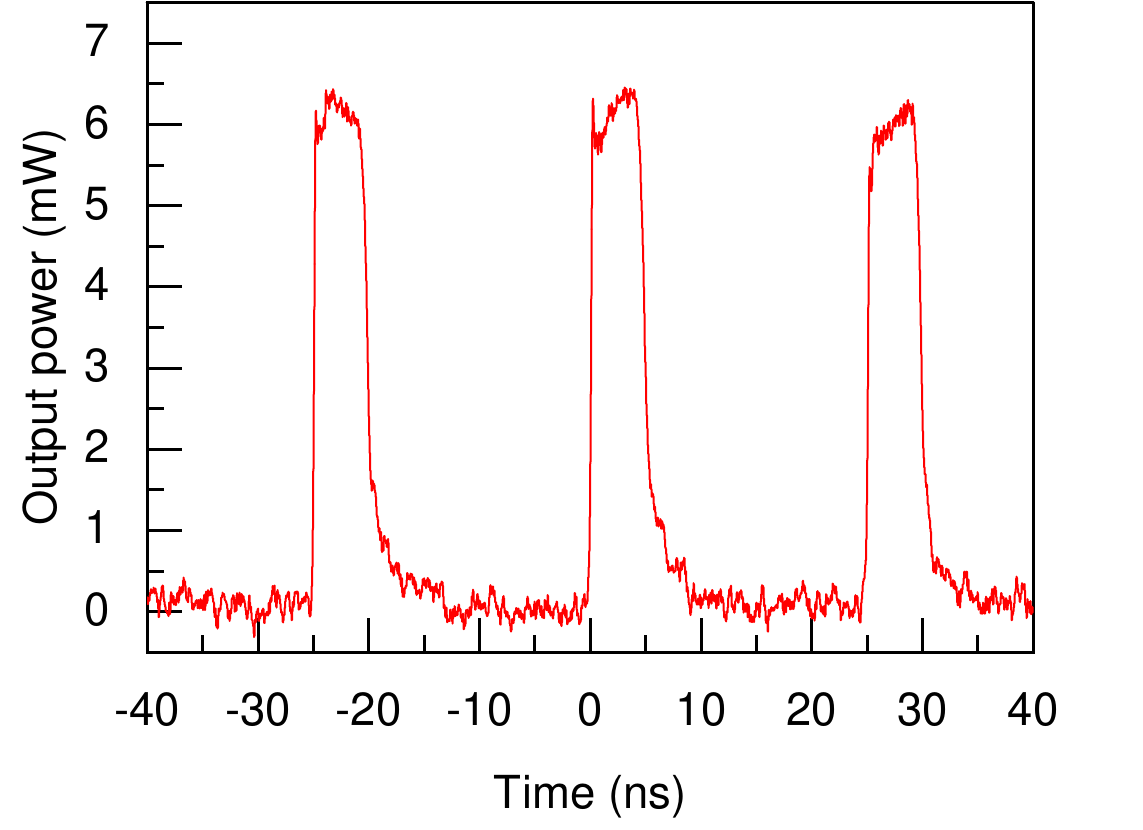}
\caption{Performance of OPL at $40$-$\mega\hertz$ pulsed light. The length of the acrylic prism is $50.8~\milli\meter$, and the average optical power of the pulsed light is $10~\milli\watt$.}
\label{fig:508-10}
\end{figure}

For the optical pulses with repetition rate at $40~\mega\hertz$, the peak power of the transmitted optical pulses can reach from several milliwatts to $38.83~\milli\watt$ as maximum. Our testing shows that optical pulses transmitted through the OPL may be sufficient to perform a pulse illumination attack on the APDs as shown in Ref.~\onlinecite{wu2020}. Specifically, under the average input power of $10~\milli\watt$, the sequence of optical pulses transmitted through the OPL with $50.8$-$\milli\meter$-long acrylic prism is shown in~\cref{fig:508-10}, presenting the peak power above $6~\milli\watt$. This amount of peak power with about $5$-$\nano\second$ pulse width provides similar pulse shape to that of the blinding pulses used in the pulse illumination attack, which indicates that there is enough energy of each optical pulses to blind the APD tested in Ref.~\onlinecite{wu2020}. Moreover, the repetition rate tested in this study, $40$-$\mega\hertz$, is also as the same as that of the pulse illumination attack demonstrated in Ref.~\onlinecite{wu2020}. Thus, the OPL equipped with such an acrylic prism may not be able to prevent a QKD system from the pulse illumination attack. It is notable that the OPL with a longer acrylic prism, $101.6~\milli\meter$, attenuates the peak power further to mitigate the threat of the pulse illumination attack.

Optical pulses with a repetition rate of $1$-$\giga\hertz$ and $40$-$\mega\hertz$ can support Eve's Trojan-horse attack on a QKD system, especially for a high-speed one. As shown by the test in~\cref{subsec:1GHz}, the optical pulses with $1$-$\giga\hertz$ repetition frequency inject into the OPL, resulting transmitted peak power higher than that of the c.w.\ light and the $40$-$\mega\hertz$ optical pulses. For a QKD system also working at $1$-$\giga\hertz$ repetition rate, these high-speed optical pulses transmitted the OPL may reach the quantum-state modulators, like phase modulators, intensity modulators, or polarization modulators, and then to be reflected. The reflected optical pulses carry the modulation information of the prepared quantum states. Thus, Eve can obtain the information about the secret key via the injected and reflected Trojan-horse pulses. 

\begin{figure}[htbp]
\centering
\includegraphics[width=0.5\textwidth]{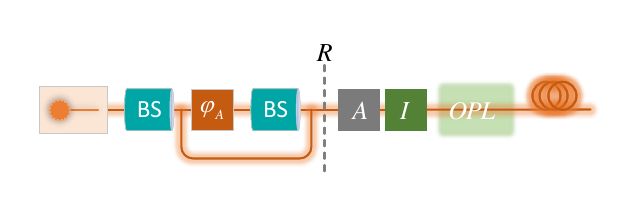}
\caption{Reflected signal model for QKD with Trojan-horse attack. $\varphi _A$ is the encoding device. $R$ is total reflection from all components to the left of the dot-dashed line; $A$ is an attenuator; $I$ is an optical isolator. }
\label{fig:tha}
\end{figure}

Previous study has shown the passive architecture against Trojan-horse attack (schematically  shown in~\cref{fig:tha}) when the average photon number of the reflected light is sufficiently low~\cite{lucamarini2015}. As can be seen from~\cref{fig:tha}, the Trojan-horse light enters via quantum channel at the right side, passes through the OPL, an optical isolator $I$, an optical attenuator $A$, and then is reflected at the internal optical components ($R$ is the total reflection of the internal optical elements to the left of the dot-dashed line), which finally returns along the original path. Therefore, the optical isolation against Trojan-horse attack can be calculated as, $\gamma = O^2 \times I^n \times A^2 \times R$, where $n$ represents the number of optical isolators present in the system. Under normal circumstances, the attenuation value of the OPL itself increases with the enhancement of the injected light. However, our experimental results show that Eve's hacking makes the attenuation value of OPL decrease. Therefore, the optical isolation of the OPL itself becomes untrusted from the security point of view. To be security, the optical isolation against Trojan-horse attack can be further written as

\begin{equation}
  \gamma = I^n \times A^2 \times R.
  \label{eq:gamma}
\end{equation}
Converting the absolute optical isolation value to a decibel ($\deci\bel$) value can more intuitively reflect the attenuation of the component. If the absolute value of the optical isolation of the component is $x$, we use the following notation, $\dot{x} = 10\log_{10}x$~\cite{lucamarini2015}. \Cref{eq:gamma} can be conveniently rewritten in dB,

\begin{equation}
  \dot{\gamma} = n\dot{I} + 2\dot{A} + \dot{R}.
\end{equation}
It should be noted that the passive architecture in~\cref{fig:tha} guarantees security against Trojan-horse attack only when the average photon leakage number should be $\mu_{\rm{out}}=10^{-6}$ ($\dot{\mu}_{\rm{out}}=-60~\deci\bel$). That is, we can guarantee the security against Trojan-horse attack by reasonably controlling the isolation of the components, i.e.,

\begin{equation}
  \dot{\mu}_{\rm{out}}=\dot{\gamma} +\dot{\chi}.
  \label{equ:mu-gamma}
\end{equation}

\Cref{fig:pulse_40mhz} shows that when the system frequency is $40$-$\mega\hertz$, the maximum number of photons passing through is $\chi=1.39113 \times 10^9$ ($\dot{\chi} \simeq 90~\deci\bel$). From \cref{equ:mu-gamma}, we then get $\dot{\gamma} = \dot{\mu}_{\rm{out}} - \dot{\chi} = (-60-90)~\deci\bel = -150~\deci\bel$. \Cref{fig:pulse_1ghz} shows that when the system frequency is $1$-$\giga\hertz$, the maximum number of photons passing through is $\chi=0.7461649 \times 10^8$ ($\dot{\chi}\simeq 80~\deci\bel$). Similarly, we then get $\dot{\gamma}=-140~\deci\bel$. This result is the total optical isolation required for security in Alice's module. Based on this, we provide several combinations of attenuation values for other components after using OPL, so that the system can prevent Trojan-horse attack. For convenience, we report the absolute value of the component, as shown in~\cref{table:tha} that contains some possible combinations.

\begin{table}[htbp]
  \centering
  \caption{Practical combinations of system components to passive architecture against Trojan-horse attack. All dotted quantities are in decibels and are given in absolute value.
  }
  \begin{ruledtabular}
  \begin{tabular}{ccccc}
    Clock rate  & $\left| {\dot \gamma } \right|$ &$\left| {\dot R } \right|$ & $\left| {\dot A } \right|$ & $\left| {\dot I } \right|$\\
    \colrule
    $1$-$\giga\hertz$ &  140 & 40 & 40 & 60(1) \\
    $1~\giga\hertz$&  140 & 20 & 0 & 60(2) \\
    $40$-$\mega\hertz$ &  150 & 40 & 10 & 50(2) \\
    $40$-$\mega\hertz$ &  150 & 30 & 0 & 60(2) \\
  \end{tabular}
  \end{ruledtabular}
  \label{table:tha}
\end{table}

\subsection{Security boundary of the blinding attack under the protection of the OPL}

\begin{figure}[ht]
\centering
\includegraphics[width=0.5\textwidth]{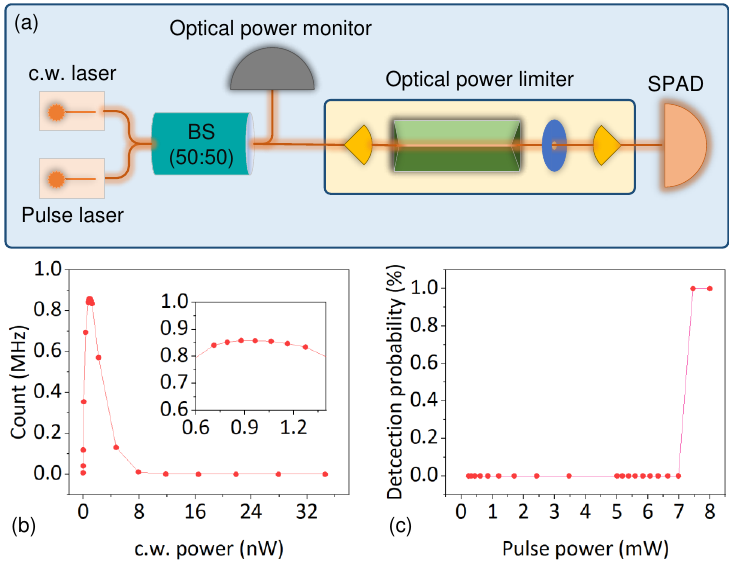}
\caption{(a) The testing scheme of the blinding attack on SPAD under the protection of the OPL. Eve applies a c.w.\ laser to blind the SPAD, which is superposed by a pulsed laser to trigger the click of SPAD. The power of these injected lasers are monitored by an optical power monitor. The length of the acrylic prism is $50.8~\milli\meter$, and the diaphragm width of $800~\micro\meter$. The model number of the SPAD is ID210, which operates in external triggering mode with the detection efficiency of 10\%, the gate width of  $4.99~\nano\second$, the gated frequency of $125~\mega\hertz$, and the dead time of $100~\nano\second$. (b) The experimental graph shows the relationship between the power of  c.w.\ blinding light and the count rate of the SPAD. (c) The detection probability of the SPAD as a function of the trigger-pulse energy. The trigger pulses are generated with the repetition frequency of $125~\mega\hertz$ and a full width at half maximum of $2.2~\nano\second$.}
\label{fig:blindatk}
\end{figure}

Some attacks may use both c.w.\ light and pulsed light, such as blinding attack on SPAD in Bob. Therefore, the security boundary under the protection of the OPL depends on both its responses of c.w.\ and pulsed injection light. To further investigate this mixed case, we demonstrate an experiment of blinding attack on a SPAD with protection of the OPL as a countermeasure, as depicted in Fig.~14(a). The specific steps of the experiment are as follows. First, c.w.\ light is injected into the OPL, whose results are shown in Fig.~14(b). When the optical power injected into the OPL reaches $34.58~\nano\watt$, the SPAD exhibits blinding effects. Then, pulsed light superposed upon the blinding light is injected into the OPL. As shown in Fig.~12(c), when the peak optical power injected into the OPL reaches $8~\milli\watt$, the detection efficiency of the SPAD becomes $100~\%$ to fully control the detection result.  Furthermore, we also conduct the same blinding-attack experiment on an OPL equipped with a $101.6$-$\milli\meter$-long acrylic prism. In this scenario, the optical power of the c.w.\ light needs to exceed $105.2~\nano\watt$ to blind the SPAD. Additionally, if Eve intends to control the SPAD's responses, the peak optical power of the pulsed light should be greater than $28.2~\milli\watt$.

These experiments demonstrates that even with the protection of the OPL, Eve still has chance to blind the SPAD and control its response. This phenomenon occurs because the required optical power injected into the OPL to blind the SPAD is much lower than the OPL's threshold of power limitation~($40~\milli\watt$). At this point, the OPL behaves as a component with a almost fixed insertion loss, as confirmed in Fig.~2. For the trigger pulse, although the OPL start limiting the injection power, the transmitted power of the optical pulses still enough to trigger the click of the blinded SPAD. These experiments show that the OPL has limited effectiveness in resisting weak light attacks, especially when the hacking optical power is less than the OPL's threshold of power limitation. Moreover, the attacks employing c.w.\ and pulsed light show as a complex scenario for the OPL as a countermeasure to be worked. This indicates that, in this mixed case, an effective countermeasure is challenging, and the security-boundary investigation and security evaluation is also complicated.

\section{Conclusion}
\label{sec:conclusion}

In this paper, we conducted comprehensive security tests on a passive OPL based on acrylic prism thermo-optical defocusing effect. The experiments utilized light sources covering a range from $0$ to $5~\watt$ of c.w.\ light, as well as pulsed light with frequencies of $0.5~\hertz$, $40~\mega\hertz$, and $1~\giga\hertz$. Through experimental verification, we disclose the security boundary of the OPL. First, it is undeniable that when the OPL is operating normally, the power of the injected hacking light can be limited below a certain value. However, when there is physical change in the OPL introduced by c.w.\ injected light, there is a window of reduced restrictions, which brings opportunities to Eve for an attack. Moreover, under the same average input optical power, the peak power of pulsed light can pass through the OPL more than that of c.w.\ light. The higher repetition rate of the injected pulses, the larger peak power. Further security analysis based on the testing results is given to present the security boundary. This study provides a more comprehensive understanding of the OPL, which allows one to properly use this device with known security boundary. The testing methodology is also applicable to other type of OPL to verify its security performance.

\acknowledgments
We thank V. Makarov, C. Wang and G. Zhang for helpful discussions. This work was funded by the
National Natural Science Foundation of China (Grants No. 61901483, No. 62371459 and No. 62061136011), the National Key Research and Development Program of China (Grant No. 2019QY0702), the Research Fund Program of State Key Laboratory of High Performance Computing (Grant No. 202001-02), Innovation Program for Quantum Science and Technology (2021ZD0300704), Aid Program for Science and Technology Innovative Research Team in Higher Educational Institutions of Hunan Province,
Key Research and Development Program of Hunan Province (2022GK2016), and Key project of Scientific Research of Hunan Provincial Education Department (22A0669).
K. Z.\ was supported by the Ministry of Science and Education of Russia (program NTI center for quantum communications) and Russian Science Foundation (grant 21-42-00040), Galician Regional Government (consolidation of Research Units: AtlantTIC), MICIN with funding from the European Union NextGenerationEU (PRTR-C17.I1) and the Galician Regional Government with own funding through the “Planes Complementarios de I+D+I con las Comunidades Aut´onomas.

\textit{Author contributions}: Q.P.\, B.G.\, and A.H.\ conducted the experiment. Q.P.\ and K. Z.\ analyzed the data. Q.P.\ and A.H.\ wrote the paper with input from all authors. A.H.\, J.W.\, and Y.G.\ supervised the project.

\def\bibsection{\medskip\begin{center}\rule{0.5\columnwidth}{.8pt}\end{center}\medskip} 

\begin{thebibliography}{54}%
  \makeatletter
  \providecommand \@ifxundefined [1]{%
   \@ifx{#1\undefined}
  }%
  \providecommand \@ifnum [1]{%
   \ifnum #1\expandafter \@firstoftwo
   \else \expandafter \@secondoftwo
   \fi
  }%
  \providecommand \@ifx [1]{%
   \ifx #1\expandafter \@firstoftwo
   \else \expandafter \@secondoftwo
   \fi
  }%
  \providecommand \natexlab [1]{#1}%
  \providecommand \enquote  [1]{``#1''}%
  \providecommand \bibnamefont  [1]{#1}%
  \providecommand \bibfnamefont [1]{#1}%
  \providecommand \citenamefont [1]{#1}%
  \providecommand \href@noop [0]{\@secondoftwo}%
  \providecommand \href [0]{\begingroup \@sanitize@url \@href}%
  \providecommand \@href[1]{\@@startlink{#1}\@@href}%
  \providecommand \@@href[1]{\endgroup#1\@@endlink}%
  \providecommand \@sanitize@url [0]{\catcode `\\12\catcode `\$12\catcode
    `\&12\catcode `\#12\catcode `\^12\catcode `\_12\catcode `\%12\relax}%
  \providecommand \@@startlink[1]{}%
  \providecommand \@@endlink[0]{}%
  \providecommand \url  [0]{\begingroup\@sanitize@url \@url }%
  \providecommand \@url [1]{\endgroup\@href {#1}{\urlprefix }}%
  \providecommand \urlprefix  [0]{URL }%
  \providecommand \Eprint [0]{\href }%
  \providecommand \doibase [0]{https://doi.org/}%
  \providecommand \selectlanguage [0]{\@gobble}%
  \providecommand \bibinfo  [0]{\@secondoftwo}%
  \providecommand \bibfield  [0]{\@secondoftwo}%
  \providecommand \translation [1]{[#1]}%
  \providecommand \BibitemOpen [0]{}%
  \providecommand \bibitemStop [0]{}%
  \providecommand \bibitemNoStop [0]{.\EOS\space}%
  \providecommand \EOS [0]{\spacefactor3000\relax}%
  \providecommand \BibitemShut  [1]{\csname bibitem#1\endcsname}%
  \let\auto@bib@innerbib\@empty
  \bibitem [{\citenamefont {Lo}\ \emph {et~al.}(2014)\citenamefont {Lo},
    \citenamefont {Curty},\ and\ \citenamefont {Tamaki}}]{lo2014}%
    \BibitemOpen
    \bibfield  {author} {\bibinfo {author} {\bibfnamefont {H.-K.}\ \bibnamefont
    {Lo}}, \bibinfo {author} {\bibfnamefont {M.}~\bibnamefont {Curty}},\ and\
    \bibinfo {author} {\bibfnamefont {K.}~\bibnamefont {Tamaki}},\ }\bibfield
    {title} {\bibinfo {title} {Secure quantum key distribution},\ }\href
    {https://doi.org/10.1038/nphoton.2014.149} {\bibfield  {journal} {\bibinfo
    {journal} {Nat. Photonics}\ }\textbf {\bibinfo {volume} {8}},\ \bibinfo
    {pages} {595} (\bibinfo {year} {2014})}\BibitemShut {NoStop}%
  \bibitem [{\citenamefont {Gisin}\ \emph {et~al.}(2002)\citenamefont {Gisin},
    \citenamefont {Ribordy}, \citenamefont {Tittel},\ and\ \citenamefont
    {Zbinden}}]{gisin2002}%
    \BibitemOpen
    \bibfield  {author} {\bibinfo {author} {\bibfnamefont {N.}~\bibnamefont
    {Gisin}}, \bibinfo {author} {\bibfnamefont {G.}~\bibnamefont {Ribordy}},
    \bibinfo {author} {\bibfnamefont {W.}~\bibnamefont {Tittel}},\ and\ \bibinfo
    {author} {\bibfnamefont {H.}~\bibnamefont {Zbinden}},\ }\bibfield  {title}
    {\bibinfo {title} {Quantum cryptography},\ }\href
    {https://doi.org/10.1103/RevModPhys.74.145} {\bibfield  {journal} {\bibinfo
    {journal} {Rev. Mod. Phys.}\ }\textbf {\bibinfo {volume} {74}},\ \bibinfo
    {pages} {145} (\bibinfo {year} {2002})}\BibitemShut {NoStop}%
  \bibitem [{\citenamefont {Diamanti}\ \emph {et~al.}(2016)\citenamefont
    {Diamanti}, \citenamefont {Lo}, \citenamefont {Qi},\ and\ \citenamefont
    {Yuan}}]{diamanti2016}%
    \BibitemOpen
    \bibfield  {author} {\bibinfo {author} {\bibfnamefont {E.}~\bibnamefont
    {Diamanti}}, \bibinfo {author} {\bibfnamefont {H.}~\bibnamefont {Lo}},
    \bibinfo {author} {\bibfnamefont {B.}~\bibnamefont {Qi}},\ and\ \bibinfo
    {author} {\bibfnamefont {Z.}~\bibnamefont {Yuan}},\ }\bibfield  {title}
    {\bibinfo {title} {Practical challenges in quantum key distribution},\ }\href
    {https://doi.org/10.1038/npjqi.2016.25} {\bibfield  {journal} {\bibinfo
    {journal} {npj Quantum Inf.}\ }\textbf {\bibinfo {volume} {2}},\ \bibinfo
    {pages} {1} (\bibinfo {year} {2016})}\BibitemShut {NoStop}%
  \bibitem [{\citenamefont {Takesue}\ \emph {et~al.}(2007)\citenamefont
    {Takesue}, \citenamefont {Nam}, \citenamefont {Zhang}, \citenamefont
    {Hadfield}, \citenamefont {Honjo}, \citenamefont {Tamaki},\ and\
    \citenamefont {Yamamoto}}]{takesue2007}%
    \BibitemOpen
    \bibfield  {author} {\bibinfo {author} {\bibfnamefont {H.}~\bibnamefont
    {Takesue}}, \bibinfo {author} {\bibfnamefont {S.~W.}\ \bibnamefont {Nam}},
    \bibinfo {author} {\bibfnamefont {Q.}~\bibnamefont {Zhang}}, \bibinfo
    {author} {\bibfnamefont {R.~H.}\ \bibnamefont {Hadfield}}, \bibinfo {author}
    {\bibfnamefont {T.}~\bibnamefont {Honjo}}, \bibinfo {author} {\bibfnamefont
    {K.}~\bibnamefont {Tamaki}},\ and\ \bibinfo {author} {\bibfnamefont
    {Y.}~\bibnamefont {Yamamoto}},\ }\bibfield  {title} {\bibinfo {title}
    {Quantum key distribution over a 40-d{B} channel loss using superconducting
    single-photon detectors},\ }\href {https://doi.org/10.1038/nphoton.2007.75}
    {\bibfield  {journal} {\bibinfo  {journal} {Nat. Photonics}\ }\textbf
    {\bibinfo {volume} {1}},\ \bibinfo {pages} {343} (\bibinfo {year}
    {2007})}\BibitemShut {NoStop}%
  \bibitem [{\citenamefont {Sibson}\ \emph {et~al.}(2017)\citenamefont {Sibson},
    \citenamefont {Kennard}, \citenamefont {Stanisic}, \citenamefont {Erven},
    \citenamefont {O’Brien},\ and\ \citenamefont {Thompson}}]{sibson2017}%
    \BibitemOpen
    \bibfield  {author} {\bibinfo {author} {\bibfnamefont {P.}~\bibnamefont
    {Sibson}}, \bibinfo {author} {\bibfnamefont {J.~E.}\ \bibnamefont {Kennard}},
    \bibinfo {author} {\bibfnamefont {S.}~\bibnamefont {Stanisic}}, \bibinfo
    {author} {\bibfnamefont {C.}~\bibnamefont {Erven}}, \bibinfo {author}
    {\bibfnamefont {J.~L.}\ \bibnamefont {O’Brien}},\ and\ \bibinfo {author}
    {\bibfnamefont {M.~G.}\ \bibnamefont {Thompson}},\ }\bibfield  {title}
    {\bibinfo {title} {Integrated silicon photonics for high-speed quantum key
    distribution},\ }\href {https://doi.org/10.1364/optica.4.000172} {\bibfield
    {journal} {\bibinfo  {journal} {Optica}\ }\textbf {\bibinfo {volume} {4}},\
    \bibinfo {pages} {172} (\bibinfo {year} {2017})}\BibitemShut {NoStop}%
  \bibitem [{\citenamefont {Eriksson}\ \emph {et~al.}(2019)\citenamefont
    {Eriksson}, \citenamefont {Hirano}, \citenamefont {Puttnam}, \citenamefont
    {Rademacher}, \citenamefont {Lu{\'\i}s}, \citenamefont {Fujiwara},
    \citenamefont {Namiki}, \citenamefont {Awaji}, \citenamefont {Takeoka},\ and\
    \citenamefont {Wada}}]{eriksson2019}%
    \BibitemOpen
    \bibfield  {author} {\bibinfo {author} {\bibfnamefont {T.~A.}\ \bibnamefont
    {Eriksson}}, \bibinfo {author} {\bibfnamefont {T.}~\bibnamefont {Hirano}},
    \bibinfo {author} {\bibfnamefont {B.~J.}\ \bibnamefont {Puttnam}}, \bibinfo
    {author} {\bibfnamefont {G.}~\bibnamefont {Rademacher}}, \bibinfo {author}
    {\bibfnamefont {R.~S.}\ \bibnamefont {Lu{\'\i}s}}, \bibinfo {author}
    {\bibfnamefont {M.}~\bibnamefont {Fujiwara}}, \bibinfo {author}
    {\bibfnamefont {R.}~\bibnamefont {Namiki}}, \bibinfo {author} {\bibfnamefont
    {Y.}~\bibnamefont {Awaji}}, \bibinfo {author} {\bibfnamefont
    {M.}~\bibnamefont {Takeoka}},\ and\ \bibinfo {author} {\bibfnamefont
    {N.}~\bibnamefont {Wada}},\ }\bibfield  {title} {\bibinfo {title} {Wavelength
    division multiplexing of continuous variable quantum key distribution and
    18.3 tbit/s data channels},\ }\href
    {https://doi.org/10.1038/s42005-018-0105-5} {\bibfield  {journal} {\bibinfo
    {journal} {Communications Physics}\ }\textbf {\bibinfo {volume} {2}},\
    \bibinfo {pages} {1} (\bibinfo {year} {2019})}\BibitemShut {NoStop}%
  \bibitem [{\citenamefont {L\"utkenhaus}(2000)}]{lutkenhaus2000}%
    \BibitemOpen
    \bibfield  {author} {\bibinfo {author} {\bibfnamefont {N.}~\bibnamefont
    {L\"utkenhaus}},\ }\bibfield  {title} {\bibinfo {title} {Security against
    individual attacks for realistic quantum key distribution},\ }\href
    {https://doi.org/10.1103/PhysRevA.61.052304} {\bibfield  {journal} {\bibinfo
    {journal} {Phys. Rev. A}\ }\textbf {\bibinfo {volume} {61}},\ \bibinfo
    {pages} {052304} (\bibinfo {year} {2000})}\BibitemShut {NoStop}%
  \bibitem [{\citenamefont {Makarov}\ \emph {et~al.}(2006)\citenamefont
    {Makarov}, \citenamefont {Anisimov},\ and\ \citenamefont
    {Skaar}}]{makarov2006}%
    \BibitemOpen
    \bibfield  {author} {\bibinfo {author} {\bibfnamefont {V.}~\bibnamefont
    {Makarov}}, \bibinfo {author} {\bibfnamefont {A.}~\bibnamefont {Anisimov}},\
    and\ \bibinfo {author} {\bibfnamefont {J.}~\bibnamefont {Skaar}},\ }\bibfield
     {title} {\bibinfo {title} {Effects of detector efficiency mismatch on
    security of quantum cryptosystems},\ }\href
    {https://doi.org/10.1103/PhysRevA.74.022313} {\bibfield  {journal} {\bibinfo
    {journal} {Phys. Rev. A}\ }\textbf {\bibinfo {volume} {74}},\ \bibinfo
    {pages} {022313} (\bibinfo {year} {2006})},\ \bibinfo {note} {erratum ibid.
    \textbf{78}, 019905 (2008)}\BibitemShut {NoStop}%
  \bibitem [{\citenamefont {Fung}\ \emph {et~al.}(2007)\citenamefont {Fung},
    \citenamefont {Qi}, \citenamefont {Tamaki},\ and\ \citenamefont
    {Lo}}]{fung2007}%
    \BibitemOpen
    \bibfield  {author} {\bibinfo {author} {\bibfnamefont {C.~H.~F.}\
    \bibnamefont {Fung}}, \bibinfo {author} {\bibfnamefont {B.}~\bibnamefont
    {Qi}}, \bibinfo {author} {\bibfnamefont {K.}~\bibnamefont {Tamaki}},\ and\
    \bibinfo {author} {\bibfnamefont {H.~K.}\ \bibnamefont {Lo}},\ }\bibfield
    {title} {\bibinfo {title} {Phase-remapping attack in practical
    quantum-key-distribution systems},\ }\href
    {https://doi.org/10.1103/PhysRevA.75.032314} {\bibfield  {journal} {\bibinfo
    {journal} {Phys. Rev. A}\ }\textbf {\bibinfo {volume} {75}},\ \bibinfo
    {pages} {032314} (\bibinfo {year} {2007})}\BibitemShut {NoStop}%
  \bibitem [{\citenamefont {Qi}\ \emph {et~al.}(2007)\citenamefont {Qi},
    \citenamefont {Fung}, \citenamefont {Lo},\ and\ \citenamefont {Ma}}]{qi2007}%
    \BibitemOpen
    \bibfield  {author} {\bibinfo {author} {\bibfnamefont {B.}~\bibnamefont
    {Qi}}, \bibinfo {author} {\bibfnamefont {C.-H.~F.}\ \bibnamefont {Fung}},
    \bibinfo {author} {\bibfnamefont {H.-K.}\ \bibnamefont {Lo}},\ and\ \bibinfo
    {author} {\bibfnamefont {X.}~\bibnamefont {Ma}},\ }\bibfield  {title}
    {\bibinfo {title} {Time-shift attack in practical quantum cryptosystems},\
    }\href@noop {} {\bibfield  {journal} {\bibinfo  {journal} {Quantum Inf.
    Comput.}\ }\textbf {\bibinfo {volume} {7}},\ \bibinfo {pages} {73} (\bibinfo
    {year} {2007})}\BibitemShut {NoStop}%
  \bibitem [{\citenamefont {Lamas-Linares}\ and\ \citenamefont
    {Kurtsiefer}(2007)}]{lamas-linares2007}%
    \BibitemOpen
    \bibfield  {author} {\bibinfo {author} {\bibfnamefont {A.}~\bibnamefont
    {Lamas-Linares}}\ and\ \bibinfo {author} {\bibfnamefont {C.}~\bibnamefont
    {Kurtsiefer}},\ }\bibfield  {title} {\bibinfo {title} {Breaking a quantum key
    distribution system through a timing side channel},\ }\href
    {https://doi.org/10.1364/oe.15.009388} {\bibfield  {journal} {\bibinfo
    {journal} {Opt. Express}\ }\textbf {\bibinfo {volume} {15}},\ \bibinfo
    {pages} {9388} (\bibinfo {year} {2007})}\BibitemShut {NoStop}%
  \bibitem [{\citenamefont {Zhao}\ \emph {et~al.}(2008)\citenamefont {Zhao},
    \citenamefont {Fung}, \citenamefont {Qi}, \citenamefont {Chen},\ and\
    \citenamefont {Lo}}]{zhao2008}%
    \BibitemOpen
    \bibfield  {author} {\bibinfo {author} {\bibfnamefont {Y.}~\bibnamefont
    {Zhao}}, \bibinfo {author} {\bibfnamefont {C.~H.~F.}\ \bibnamefont {Fung}},
    \bibinfo {author} {\bibfnamefont {B.}~\bibnamefont {Qi}}, \bibinfo {author}
    {\bibfnamefont {C.}~\bibnamefont {Chen}},\ and\ \bibinfo {author}
    {\bibfnamefont {H.-K.}\ \bibnamefont {Lo}},\ }\bibfield  {title} {\bibinfo
    {title} {Quantum hacking: Experimental demonstration of time-shift attack
    against practical quantum-key-distribution systems},\ }\href
    {https://doi.org/10.1103/PhysRevA.78.042333} {\bibfield  {journal} {\bibinfo
    {journal} {Phys. Rev. A}\ }\textbf {\bibinfo {volume} {78}},\ \bibinfo
    {pages} {042333} (\bibinfo {year} {2008})}\BibitemShut {NoStop}%
  \bibitem [{\citenamefont {Lydersen}\ \emph
    {et~al.}(2010{\natexlab{a}})\citenamefont {Lydersen}, \citenamefont
    {Wiechers}, \citenamefont {Wittmann}, \citenamefont {Elser}, \citenamefont
    {Skaar},\ and\ \citenamefont {Makarov}}]{lydersen2010a}%
    \BibitemOpen
    \bibfield  {author} {\bibinfo {author} {\bibfnamefont {L.}~\bibnamefont
    {Lydersen}}, \bibinfo {author} {\bibfnamefont {C.}~\bibnamefont {Wiechers}},
    \bibinfo {author} {\bibfnamefont {C.}~\bibnamefont {Wittmann}}, \bibinfo
    {author} {\bibfnamefont {D.}~\bibnamefont {Elser}}, \bibinfo {author}
    {\bibfnamefont {J.}~\bibnamefont {Skaar}},\ and\ \bibinfo {author}
    {\bibfnamefont {V.}~\bibnamefont {Makarov}},\ }\bibfield  {title} {\bibinfo
    {title} {Hacking commercial quantum cryptography systems by tailored bright
    illumination},\ }\href {https://doi.org/10.1038/nphoton.2010.214} {\bibfield
    {journal} {\bibinfo  {journal} {Nat. Photonics}\ }\textbf {\bibinfo {volume}
    {4}},\ \bibinfo {pages} {686} (\bibinfo {year}
    {2010}{\natexlab{a}})}\BibitemShut {NoStop}%
  \bibitem [{\citenamefont {Lydersen}\ \emph
    {et~al.}(2010{\natexlab{b}})\citenamefont {Lydersen}, \citenamefont
    {Wiechers}, \citenamefont {Wittmann}, \citenamefont {Elser}, \citenamefont
    {Skaar},\ and\ \citenamefont {Makarov}}]{lydersen2010b}%
    \BibitemOpen
    \bibfield  {author} {\bibinfo {author} {\bibfnamefont {L.}~\bibnamefont
    {Lydersen}}, \bibinfo {author} {\bibfnamefont {C.}~\bibnamefont {Wiechers}},
    \bibinfo {author} {\bibfnamefont {C.}~\bibnamefont {Wittmann}}, \bibinfo
    {author} {\bibfnamefont {D.}~\bibnamefont {Elser}}, \bibinfo {author}
    {\bibfnamefont {J.}~\bibnamefont {Skaar}},\ and\ \bibinfo {author}
    {\bibfnamefont {V.}~\bibnamefont {Makarov}},\ }\bibfield  {title} {\bibinfo
    {title} {Thermal blinding of gated detectors in quantum cryptography},\
    }\href {https://doi.org/10.1364/oe.18.027938} {\bibfield  {journal} {\bibinfo
     {journal} {Opt. Express}\ }\textbf {\bibinfo {volume} {18}},\ \bibinfo
    {pages} {27938} (\bibinfo {year} {2010}{\natexlab{b}})}\BibitemShut {NoStop}%
  \bibitem [{\citenamefont {Xu}\ \emph {et~al.}(2010)\citenamefont {Xu},
    \citenamefont {Qi},\ and\ \citenamefont {Lo}}]{xu2010}%
    \BibitemOpen
    \bibfield  {author} {\bibinfo {author} {\bibfnamefont {F.}~\bibnamefont
    {Xu}}, \bibinfo {author} {\bibfnamefont {B.}~\bibnamefont {Qi}},\ and\
    \bibinfo {author} {\bibfnamefont {H.-K.}\ \bibnamefont {Lo}},\ }\bibfield
    {title} {\bibinfo {title} {Experimental demonstration of phase-remapping
    attack in a practical quantum key distribution system},\ }\href
    {https://doi.org/10.1088/1367-2630/12/11/113026} {\bibfield  {journal}
    {\bibinfo  {journal} {New J. Phys.}\ }\textbf {\bibinfo {volume} {12}},\
    \bibinfo {pages} {113026} (\bibinfo {year} {2010})}\BibitemShut {NoStop}%
  \bibitem [{\citenamefont {Li}\ \emph {et~al.}(2011)\citenamefont {Li},
    \citenamefont {Wang}, \citenamefont {Huang}, \citenamefont {Chen},
    \citenamefont {Yin}, \citenamefont {Li}, \citenamefont {Zhou}, \citenamefont
    {Liu}, \citenamefont {Zhang}, \citenamefont {Guo}, \citenamefont {Bao},\ and\
    \citenamefont {Han}}]{li2011a}%
    \BibitemOpen
    \bibfield  {author} {\bibinfo {author} {\bibfnamefont {H.-W.}\ \bibnamefont
    {Li}}, \bibinfo {author} {\bibfnamefont {S.}~\bibnamefont {Wang}}, \bibinfo
    {author} {\bibfnamefont {J.-Z.}\ \bibnamefont {Huang}}, \bibinfo {author}
    {\bibfnamefont {W.}~\bibnamefont {Chen}}, \bibinfo {author} {\bibfnamefont
    {Z.-Q.}\ \bibnamefont {Yin}}, \bibinfo {author} {\bibfnamefont {F.-Y.}\
    \bibnamefont {Li}}, \bibinfo {author} {\bibfnamefont {Z.}~\bibnamefont
    {Zhou}}, \bibinfo {author} {\bibfnamefont {D.}~\bibnamefont {Liu}}, \bibinfo
    {author} {\bibfnamefont {Y.}~\bibnamefont {Zhang}}, \bibinfo {author}
    {\bibfnamefont {G.-C.}\ \bibnamefont {Guo}}, \bibinfo {author} {\bibfnamefont
    {W.-S.}\ \bibnamefont {Bao}},\ and\ \bibinfo {author} {\bibfnamefont {Z.-F.}\
    \bibnamefont {Han}},\ }\bibfield  {title} {\bibinfo {title} {Attacking a
    practical quantum-key-distribution system with wavelength-dependent
    beam-splitter and multiwavelength sources},\ }\href
    {https://doi.org/10.1103/PhysRevA.84.062308} {\bibfield  {journal} {\bibinfo
    {journal} {Phys. Rev. A}\ }\textbf {\bibinfo {volume} {84}},\ \bibinfo
    {pages} {062308} (\bibinfo {year} {2011})}\BibitemShut {NoStop}%
  \bibitem [{\citenamefont {Wiechers}\ \emph {et~al.}(2011)\citenamefont
    {Wiechers}, \citenamefont {Lydersen}, \citenamefont {Wittmann}, \citenamefont
    {Elser}, \citenamefont {Skaar}, \citenamefont {Marquardt}, \citenamefont
    {Makarov},\ and\ \citenamefont {Leuchs}}]{wiechers2011}%
    \BibitemOpen
    \bibfield  {author} {\bibinfo {author} {\bibfnamefont {C.}~\bibnamefont
    {Wiechers}}, \bibinfo {author} {\bibfnamefont {L.}~\bibnamefont {Lydersen}},
    \bibinfo {author} {\bibfnamefont {C.}~\bibnamefont {Wittmann}}, \bibinfo
    {author} {\bibfnamefont {D.}~\bibnamefont {Elser}}, \bibinfo {author}
    {\bibfnamefont {J.}~\bibnamefont {Skaar}}, \bibinfo {author} {\bibfnamefont
    {C.}~\bibnamefont {Marquardt}}, \bibinfo {author} {\bibfnamefont
    {V.}~\bibnamefont {Makarov}},\ and\ \bibinfo {author} {\bibfnamefont
    {G.}~\bibnamefont {Leuchs}},\ }\bibfield  {title} {\bibinfo {title}
    {After-gate attack on a quantum cryptosystem},\ }\href
    {https://doi.org/10.1088/1367-2630/13/1/013043} {\bibfield  {journal}
    {\bibinfo  {journal} {New J. Phys.}\ }\textbf {\bibinfo {volume} {13}},\
    \bibinfo {pages} {013043} (\bibinfo {year} {2011})}\BibitemShut {NoStop}%
  \bibitem [{\citenamefont {Lydersen}\ \emph
    {et~al.}(2011{\natexlab{a}})\citenamefont {Lydersen}, \citenamefont
    {Akhlaghi}, \citenamefont {Majedi}, \citenamefont {Skaar},\ and\
    \citenamefont {Makarov}}]{lydersen2011c}%
    \BibitemOpen
    \bibfield  {author} {\bibinfo {author} {\bibfnamefont {L.}~\bibnamefont
    {Lydersen}}, \bibinfo {author} {\bibfnamefont {M.~K.}\ \bibnamefont
    {Akhlaghi}}, \bibinfo {author} {\bibfnamefont {A.~H.}\ \bibnamefont
    {Majedi}}, \bibinfo {author} {\bibfnamefont {J.}~\bibnamefont {Skaar}},\ and\
    \bibinfo {author} {\bibfnamefont {V.}~\bibnamefont {Makarov}},\ }\bibfield
    {title} {\bibinfo {title} {Controlling a superconducting nanowire
    single-photon detector using tailored bright illumination},\ }\href
    {https://doi.org/10.1088/1367-2630/13/11/113042} {\bibfield  {journal}
    {\bibinfo  {journal} {New J. Phys.}\ }\textbf {\bibinfo {volume} {13}},\
    \bibinfo {pages} {113042} (\bibinfo {year} {2011}{\natexlab{a}})}\BibitemShut
    {NoStop}%
  \bibitem [{\citenamefont {Lydersen}\ \emph
    {et~al.}(2011{\natexlab{b}})\citenamefont {Lydersen}, \citenamefont {Jain},
    \citenamefont {Wittmann}, \citenamefont {Mar{\o}y}, \citenamefont {Skaar},
    \citenamefont {Marquardt}, \citenamefont {Makarov},\ and\ \citenamefont
    {Leuchs}}]{lydersen2011b}%
    \BibitemOpen
    \bibfield  {author} {\bibinfo {author} {\bibfnamefont {L.}~\bibnamefont
    {Lydersen}}, \bibinfo {author} {\bibfnamefont {N.}~\bibnamefont {Jain}},
    \bibinfo {author} {\bibfnamefont {C.}~\bibnamefont {Wittmann}}, \bibinfo
    {author} {\bibfnamefont {{\O}.}~\bibnamefont {Mar{\o}y}}, \bibinfo {author}
    {\bibfnamefont {J.}~\bibnamefont {Skaar}}, \bibinfo {author} {\bibfnamefont
    {C.}~\bibnamefont {Marquardt}}, \bibinfo {author} {\bibfnamefont
    {V.}~\bibnamefont {Makarov}},\ and\ \bibinfo {author} {\bibfnamefont
    {G.}~\bibnamefont {Leuchs}},\ }\bibfield  {title} {\bibinfo {title}
    {Superlinear threshold detectors in quantum cryptography},\ }\href
    {https://doi.org/10.1103/PhysRevA.84.032320} {\bibfield  {journal} {\bibinfo
    {journal} {Phys. Rev. A}\ }\textbf {\bibinfo {volume} {84}},\ \bibinfo
    {pages} {032320} (\bibinfo {year} {2011}{\natexlab{b}})}\BibitemShut
    {NoStop}%
  \bibitem [{\citenamefont {Gerhardt}\ \emph {et~al.}(2011)\citenamefont
    {Gerhardt}, \citenamefont {Liu}, \citenamefont {Lamas-Linares}, \citenamefont
    {Skaar}, \citenamefont {Kurtsiefer},\ and\ \citenamefont
    {Makarov}}]{gerhardt2011}%
    \BibitemOpen
    \bibfield  {author} {\bibinfo {author} {\bibfnamefont {I.}~\bibnamefont
    {Gerhardt}}, \bibinfo {author} {\bibfnamefont {Q.}~\bibnamefont {Liu}},
    \bibinfo {author} {\bibfnamefont {A.}~\bibnamefont {Lamas-Linares}}, \bibinfo
    {author} {\bibfnamefont {J.}~\bibnamefont {Skaar}}, \bibinfo {author}
    {\bibfnamefont {C.}~\bibnamefont {Kurtsiefer}},\ and\ \bibinfo {author}
    {\bibfnamefont {V.}~\bibnamefont {Makarov}},\ }\bibfield  {title} {\bibinfo
    {title} {Full-field implementation of a perfect eavesdropper on a quantum
    cryptography system},\ }\href {https://doi.org/10.1038/ncomms1348} {\bibfield
     {journal} {\bibinfo  {journal} {Nat. Commun.}\ }\textbf {\bibinfo {volume}
    {2}},\ \bibinfo {pages} {349} (\bibinfo {year} {2011})}\BibitemShut {NoStop}%
  \bibitem [{\citenamefont {Sun}\ \emph {et~al.}(2011)\citenamefont {Sun},
    \citenamefont {Jiang},\ and\ \citenamefont {Liang}}]{sun2011}%
    \BibitemOpen
    \bibfield  {author} {\bibinfo {author} {\bibfnamefont {S.-H.}\ \bibnamefont
    {Sun}}, \bibinfo {author} {\bibfnamefont {M.-S.}\ \bibnamefont {Jiang}},\
    and\ \bibinfo {author} {\bibfnamefont {L.-M.}\ \bibnamefont {Liang}},\
    }\bibfield  {title} {\bibinfo {title} {Passive {F}araday-mirror attack in a
    practical two-way quantum-key-distribution system},\ }\href
    {https://doi.org/10.1103/PhysRevA.83.062331} {\bibfield  {journal} {\bibinfo
    {journal} {Phys. Rev. A}\ }\textbf {\bibinfo {volume} {83}},\ \bibinfo
    {pages} {062331} (\bibinfo {year} {2011})}\BibitemShut {NoStop}%
  \bibitem [{\citenamefont {Jain}\ \emph {et~al.}(2011)\citenamefont {Jain},
    \citenamefont {Wittmann}, \citenamefont {Lydersen}, \citenamefont {Wiechers},
    \citenamefont {Elser}, \citenamefont {Marquardt}, \citenamefont {Makarov},\
    and\ \citenamefont {Leuchs}}]{jain2011}%
    \BibitemOpen
    \bibfield  {author} {\bibinfo {author} {\bibfnamefont {N.}~\bibnamefont
    {Jain}}, \bibinfo {author} {\bibfnamefont {C.}~\bibnamefont {Wittmann}},
    \bibinfo {author} {\bibfnamefont {L.}~\bibnamefont {Lydersen}}, \bibinfo
    {author} {\bibfnamefont {C.}~\bibnamefont {Wiechers}}, \bibinfo {author}
    {\bibfnamefont {D.}~\bibnamefont {Elser}}, \bibinfo {author} {\bibfnamefont
    {C.}~\bibnamefont {Marquardt}}, \bibinfo {author} {\bibfnamefont
    {V.}~\bibnamefont {Makarov}},\ and\ \bibinfo {author} {\bibfnamefont
    {G.}~\bibnamefont {Leuchs}},\ }\bibfield  {title} {\bibinfo {title} {Device
    calibration impacts security of quantum key distribution},\ }\href
    {https://doi.org/10.1103/PhysRevLett.107.110501} {\bibfield  {journal}
    {\bibinfo  {journal} {Phys. Rev. Lett.}\ }\textbf {\bibinfo {volume} {107}},\
    \bibinfo {pages} {110501} (\bibinfo {year} {2011})}\BibitemShut {NoStop}%
  \bibitem [{\citenamefont {Bugge}\ \emph {et~al.}(2014)\citenamefont {Bugge},
    \citenamefont {Sauge}, \citenamefont {Ghazali}, \citenamefont {Skaar},
    \citenamefont {Lydersen},\ and\ \citenamefont {Makarov}}]{bugge2014}%
    \BibitemOpen
    \bibfield  {author} {\bibinfo {author} {\bibfnamefont {A.~N.}\ \bibnamefont
    {Bugge}}, \bibinfo {author} {\bibfnamefont {S.}~\bibnamefont {Sauge}},
    \bibinfo {author} {\bibfnamefont {AinaMardhiyahM}~\bibnamefont {Ghazali}},
    \bibinfo {author} {\bibfnamefont {J.}~\bibnamefont {Skaar}}, \bibinfo
    {author} {\bibfnamefont {L.}~\bibnamefont {Lydersen}},\ and\ \bibinfo
    {author} {\bibfnamefont {V.}~\bibnamefont {Makarov}},\ }\bibfield  {title}
    {\bibinfo {title} {Laser damage helps the eavesdropper in quantum
    cryptography},\ }\href {https://doi.org/10.1103/PhysRevLett.112.070503}
    {\bibfield  {journal} {\bibinfo  {journal} {Phys. Rev. Lett.}\ }\textbf
    {\bibinfo {volume} {112}},\ \bibinfo {pages} {070503} (\bibinfo {year}
    {2014})}\BibitemShut {NoStop}%
  \bibitem [{\citenamefont {Sajeed}\ \emph {et~al.}(2015)\citenamefont {Sajeed},
    \citenamefont {Chaiwongkhot}, \citenamefont {Bourgoin}, \citenamefont
    {Jennewein}, \citenamefont {L{\" u}tkenhaus},\ and\ \citenamefont
    {Makarov}}]{sajeed2015a}%
    \BibitemOpen
    \bibfield  {author} {\bibinfo {author} {\bibfnamefont {S.}~\bibnamefont
    {Sajeed}}, \bibinfo {author} {\bibfnamefont {P.}~\bibnamefont
    {Chaiwongkhot}}, \bibinfo {author} {\bibfnamefont {J.-P.}\ \bibnamefont
    {Bourgoin}}, \bibinfo {author} {\bibfnamefont {T.}~\bibnamefont {Jennewein}},
    \bibinfo {author} {\bibfnamefont {N.}~\bibnamefont {L{\" u}tkenhaus}},\ and\
    \bibinfo {author} {\bibfnamefont {V.}~\bibnamefont {Makarov}},\ }\bibfield
    {title} {\bibinfo {title} {Security loophole in free-space quantum key
    distribution due to spatial-mode detector-efficiency mismatch},\ }\href
    {https://doi.org/10.1103/PhysRevA.91.062301} {\bibfield  {journal} {\bibinfo
    {journal} {Phys. Rev. A}\ }\textbf {\bibinfo {volume} {91}},\ \bibinfo
    {pages} {062301} (\bibinfo {year} {2015})}\BibitemShut {NoStop}%
  \bibitem [{\citenamefont {Huang}\ \emph {et~al.}(2016)\citenamefont {Huang},
    \citenamefont {Sajeed}, \citenamefont {Chaiwongkhot}, \citenamefont
    {Soucarros}, \citenamefont {Legr{\' e}},\ and\ \citenamefont
    {Makarov}}]{huang2016}%
    \BibitemOpen
    \bibfield  {author} {\bibinfo {author} {\bibfnamefont {A.}~\bibnamefont
    {Huang}}, \bibinfo {author} {\bibfnamefont {S.}~\bibnamefont {Sajeed}},
    \bibinfo {author} {\bibfnamefont {P.}~\bibnamefont {Chaiwongkhot}}, \bibinfo
    {author} {\bibfnamefont {M.}~\bibnamefont {Soucarros}}, \bibinfo {author}
    {\bibfnamefont {M.}~\bibnamefont {Legr{\' e}}},\ and\ \bibinfo {author}
    {\bibfnamefont {V.}~\bibnamefont {Makarov}},\ }\bibfield  {title} {\bibinfo
    {title} {Testing random-detector-efficiency countermeasure in a commercial
    system reveals a breakable unrealistic assumption},\ }\href
    {https://doi.org/10.1109/JQE.2016.2611443} {\bibfield  {journal} {\bibinfo
    {journal} {IEEE J. Quantum Electron.}\ }\textbf {\bibinfo {volume} {52}},\
    \bibinfo {pages} {8000211} (\bibinfo {year} {2016})}\BibitemShut {NoStop}%
  \bibitem [{\citenamefont {Makarov}\ \emph {et~al.}(2016)\citenamefont
    {Makarov}, \citenamefont {Bourgoin}, \citenamefont {Chaiwongkhot},
    \citenamefont {Gagn{\'e}}, \citenamefont {Jennewein}, \citenamefont {Kaiser},
    \citenamefont {Kashyap}, \citenamefont {Legr{\'e}}, \citenamefont
    {Minshull},\ and\ \citenamefont {Sajeed}}]{makarov2016}%
    \BibitemOpen
    \bibfield  {author} {\bibinfo {author} {\bibfnamefont {V.}~\bibnamefont
    {Makarov}}, \bibinfo {author} {\bibfnamefont {J.-P.}\ \bibnamefont
    {Bourgoin}}, \bibinfo {author} {\bibfnamefont {P.}~\bibnamefont
    {Chaiwongkhot}}, \bibinfo {author} {\bibfnamefont {M.}~\bibnamefont
    {Gagn{\'e}}}, \bibinfo {author} {\bibfnamefont {T.}~\bibnamefont
    {Jennewein}}, \bibinfo {author} {\bibfnamefont {S.}~\bibnamefont {Kaiser}},
    \bibinfo {author} {\bibfnamefont {R.}~\bibnamefont {Kashyap}}, \bibinfo
    {author} {\bibfnamefont {M.}~\bibnamefont {Legr{\'e}}}, \bibinfo {author}
    {\bibfnamefont {C.}~\bibnamefont {Minshull}},\ and\ \bibinfo {author}
    {\bibfnamefont {S.}~\bibnamefont {Sajeed}},\ }\bibfield  {title} {\bibinfo
    {title} {Creation of backdoors in quantum communications via laser damage},\
    }\href {https://doi.org/10.1103/PhysRevA.94.030302} {\bibfield  {journal}
    {\bibinfo  {journal} {Phys. Rev. A}\ }\textbf {\bibinfo {volume} {94}},\
    \bibinfo {pages} {030302(R)} (\bibinfo {year} {2016})}\BibitemShut {NoStop}%
  \bibitem [{\citenamefont {Huang}\ \emph {et~al.}(2018)\citenamefont {Huang},
    \citenamefont {Sun}, \citenamefont {Liu},\ and\ \citenamefont
    {Makarov}}]{huang2018}%
    \BibitemOpen
    \bibfield  {author} {\bibinfo {author} {\bibfnamefont {A.}~\bibnamefont
    {Huang}}, \bibinfo {author} {\bibfnamefont {S.-H.}\ \bibnamefont {Sun}},
    \bibinfo {author} {\bibfnamefont {Z.}~\bibnamefont {Liu}},\ and\ \bibinfo
    {author} {\bibfnamefont {V.}~\bibnamefont {Makarov}},\ }\bibfield  {title}
    {\bibinfo {title} {Quantum key distribution with distinguishable decoy
    states},\ }\href {https://doi.org/10.1103/PhysRevA.98.012330} {\bibfield
    {journal} {\bibinfo  {journal} {Phys. Rev. A}\ }\textbf {\bibinfo {volume}
    {98}},\ \bibinfo {pages} {012330} (\bibinfo {year} {2018})}\BibitemShut
    {NoStop}%
  \bibitem [{\citenamefont {Qian}\ \emph {et~al.}(2018)\citenamefont {Qian},
    \citenamefont {He}, \citenamefont {Wang}, \citenamefont {Chen}, \citenamefont
    {Yin}, \citenamefont {Guo},\ and\ \citenamefont {Han}}]{qian2018}%
    \BibitemOpen
    \bibfield  {author} {\bibinfo {author} {\bibfnamefont {Y.-J.}\ \bibnamefont
    {Qian}}, \bibinfo {author} {\bibfnamefont {D.-Y.}\ \bibnamefont {He}},
    \bibinfo {author} {\bibfnamefont {S.}~\bibnamefont {Wang}}, \bibinfo {author}
    {\bibfnamefont {W.}~\bibnamefont {Chen}}, \bibinfo {author} {\bibfnamefont
    {Z.-Q.}\ \bibnamefont {Yin}}, \bibinfo {author} {\bibfnamefont {G.-C.}\
    \bibnamefont {Guo}},\ and\ \bibinfo {author} {\bibfnamefont {Z.-F.}\
    \bibnamefont {Han}},\ }\bibfield  {title} {\bibinfo {title} {Hacking the
    quantum key distribution system by exploiting the avalanche-transition region
    of single-photon detectors},\ }\href
    {https://doi.org/10.1103/PhysRevApplied.10.064062} {\bibfield  {journal}
    {\bibinfo  {journal} {Phys. Rev. Appl.}\ }\textbf {\bibinfo {volume} {10}},\
    \bibinfo {pages} {064062} (\bibinfo {year} {2018})}\BibitemShut {NoStop}%
  \bibitem [{\citenamefont {Huang}\ \emph {et~al.}(2019)\citenamefont {Huang},
    \citenamefont {Navarrete}, \citenamefont {Sun}, \citenamefont {Chaiwongkhot},
    \citenamefont {Curty},\ and\ \citenamefont {Makarov}}]{huang2019}%
    \BibitemOpen
    \bibfield  {author} {\bibinfo {author} {\bibfnamefont {A.}~\bibnamefont
    {Huang}}, \bibinfo {author} {\bibfnamefont {{\'A}.}~\bibnamefont
    {Navarrete}}, \bibinfo {author} {\bibfnamefont {S.-H.}\ \bibnamefont {Sun}},
    \bibinfo {author} {\bibfnamefont {P.}~\bibnamefont {Chaiwongkhot}}, \bibinfo
    {author} {\bibfnamefont {M.}~\bibnamefont {Curty}},\ and\ \bibinfo {author}
    {\bibfnamefont {V.}~\bibnamefont {Makarov}},\ }\bibfield  {title} {\bibinfo
    {title} {Laser-seeding attack in quantum key distribution},\ }\href
    {https://doi.org/10.1103/PhysRevApplied.12.064043} {\bibfield  {journal}
    {\bibinfo  {journal} {Phys. Rev. Appl.}\ }\textbf {\bibinfo {volume} {12}},\
    \bibinfo {pages} {064043} (\bibinfo {year} {2019})}\BibitemShut {NoStop}%
  \bibitem [{\citenamefont {Huang}\ \emph {et~al.}(2020)\citenamefont {Huang},
    \citenamefont {Li}, \citenamefont {Egorov}, \citenamefont {Tchouragoulov},
    \citenamefont {Kumar},\ and\ \citenamefont {Makarov}}]{huang2020}%
    \BibitemOpen
    \bibfield  {author} {\bibinfo {author} {\bibfnamefont {A.}~\bibnamefont
    {Huang}}, \bibinfo {author} {\bibfnamefont {R.}~\bibnamefont {Li}}, \bibinfo
    {author} {\bibfnamefont {V.}~\bibnamefont {Egorov}}, \bibinfo {author}
    {\bibfnamefont {S.}~\bibnamefont {Tchouragoulov}}, \bibinfo {author}
    {\bibfnamefont {K.}~\bibnamefont {Kumar}},\ and\ \bibinfo {author}
    {\bibfnamefont {V.}~\bibnamefont {Makarov}},\ }\bibfield  {title} {\bibinfo
    {title} {Laser damage attack against optical attenuators in quantum key
    distribution},\ }\href {https://doi.org/10.1103/PhysRevApplied.13.034017}
    {\bibfield  {journal} {\bibinfo  {journal} {Phys. Rev. Appl.}\ }\textbf
    {\bibinfo {volume} {13}},\ \bibinfo {pages} {034017} (\bibinfo {year}
    {2020})}\BibitemShut {NoStop}%
  \bibitem [{\citenamefont {Sun}\ and\ \citenamefont {Huang}(2022)}]{sun2022}%
    \BibitemOpen
    \bibfield  {author} {\bibinfo {author} {\bibfnamefont {S.}~\bibnamefont
    {Sun}}\ and\ \bibinfo {author} {\bibfnamefont {A.}~\bibnamefont {Huang}},\
    }\bibfield  {title} {\bibinfo {title} {A review of security evaluation of
    practical quantum key distribution system},\ }\href
    {https://doi.org/10.3390/e24020260} {\bibfield  {journal} {\bibinfo
    {journal} {Entropy}\ }\textbf {\bibinfo {volume} {24}},\ \bibinfo {pages}
    {260} (\bibinfo {year} {2022})}\BibitemShut {NoStop}%
  \bibitem [{\citenamefont {Chaiwongkhot}\ \emph {et~al.}(2022)\citenamefont
    {Chaiwongkhot}, \citenamefont {Zhong}, \citenamefont {Huang}, \citenamefont
    {Qin}, \citenamefont {Shi},\ and\ \citenamefont
    {Makarov}}]{chaiwongkhot2022}%
    \BibitemOpen
    \bibfield  {author} {\bibinfo {author} {\bibfnamefont {P.}~\bibnamefont
    {Chaiwongkhot}}, \bibinfo {author} {\bibfnamefont {J.}~\bibnamefont {Zhong}},
    \bibinfo {author} {\bibfnamefont {A.}~\bibnamefont {Huang}}, \bibinfo
    {author} {\bibfnamefont {H.}~\bibnamefont {Qin}}, \bibinfo {author}
    {\bibfnamefont {S.-c.}\ \bibnamefont {Shi}},\ and\ \bibinfo {author}
    {\bibfnamefont {V.}~\bibnamefont {Makarov}},\ }\bibfield  {title} {\bibinfo
    {title} {Faking photon number on a transition-edge sensor},\ }\href
    {https://doi.org/10.1140/epjqt/s40507-022-00141-2} {\bibfield  {journal}
    {\bibinfo  {journal} {EPJ Quantum Technol.}\ }\textbf {\bibinfo {volume}
    {9}},\ \bibinfo {pages} {23} (\bibinfo {year} {2022})}\BibitemShut {NoStop}%
  \bibitem [{\citenamefont {Huang}\ \emph {et~al.}()\citenamefont {Huang},
    \citenamefont {Mizutani}, \citenamefont {Lo}, \citenamefont {Makarov},\ and\
    \citenamefont {Tamaki}}]{huang2022}%
    \BibitemOpen
    \bibfield  {author} {\bibinfo {author} {\bibfnamefont {A.}~\bibnamefont
    {Huang}}, \bibinfo {author} {\bibfnamefont {A.}~\bibnamefont {Mizutani}},
    \bibinfo {author} {\bibfnamefont {H.-K.}\ \bibnamefont {Lo}}, \bibinfo
    {author} {\bibfnamefont {V.}~\bibnamefont {Makarov}},\ and\ \bibinfo {author}
    {\bibfnamefont {K.}~\bibnamefont {Tamaki}},\ }\bibfield  {title} {\bibinfo
    {title} {Characterisation of state preparation uncertainty in quantum key
    distribution},\ }\href@noop {} {\bibfield  {journal} {\bibinfo  {journal}
    {arXiv}\ }}\Eprint {https://arxiv.org/abs/2205.11870} {arXiv:2205.11870
    [quant-ph]} \BibitemShut {NoStop}%
  \bibitem [{\citenamefont {Gao}\ \emph {et~al.}(2022)\citenamefont {Gao},
    \citenamefont {Wu}, \citenamefont {Shi}, \citenamefont {Liu}, \citenamefont
    {Wang}, \citenamefont {Yu}, \citenamefont {Huang},\ and\ \citenamefont
    {Wu}}]{gao2022}%
    \BibitemOpen
    \bibfield  {author} {\bibinfo {author} {\bibfnamefont {B.}~\bibnamefont
    {Gao}}, \bibinfo {author} {\bibfnamefont {Z.}~\bibnamefont {Wu}}, \bibinfo
    {author} {\bibfnamefont {W.}~\bibnamefont {Shi}}, \bibinfo {author}
    {\bibfnamefont {Y.}~\bibnamefont {Liu}}, \bibinfo {author} {\bibfnamefont
    {D.}~\bibnamefont {Wang}}, \bibinfo {author} {\bibfnamefont {C.}~\bibnamefont
    {Yu}}, \bibinfo {author} {\bibfnamefont {A.}~\bibnamefont {Huang}},\ and\
    \bibinfo {author} {\bibfnamefont {J.}~\bibnamefont {Wu}},\ }\bibfield
    {title} {\bibinfo {title} {Strong pulse illumination hacks self-differencing
    avalanche photodiode detectors in a high-speed quantum key distribution
    system},\ }\href {https://doi.org/10.1103/PhysRevA.106.033713} {\bibfield
    {journal} {\bibinfo  {journal} {Phys. Rev. A}\ }\textbf {\bibinfo {volume}
    {106}},\ \bibinfo {pages} {033713} (\bibinfo {year} {2022})}\BibitemShut
    {NoStop}%
  \bibitem [{\citenamefont {Zheng}\ \emph {et~al.}(2019)\citenamefont {Zheng},
    \citenamefont {Huang}, \citenamefont {Huang}, \citenamefont {Peng},\ and\
    \citenamefont {Zeng}}]{zheng2019}%
    \BibitemOpen
    \bibfield  {author} {\bibinfo {author} {\bibfnamefont {Y.}~\bibnamefont
    {Zheng}}, \bibinfo {author} {\bibfnamefont {P.}~\bibnamefont {Huang}},
    \bibinfo {author} {\bibfnamefont {A.}~\bibnamefont {Huang}}, \bibinfo
    {author} {\bibfnamefont {J.}~\bibnamefont {Peng}},\ and\ \bibinfo {author}
    {\bibfnamefont {G.}~\bibnamefont {Zeng}},\ }\bibfield  {title} {\bibinfo
    {title} {Security analysis of practical continuous-variable quantum key
    distribution systems under laser seeding attack},\ }\href
    {https://doi.org/10.1364/oe.27.027369} {\bibfield  {journal} {\bibinfo
    {journal} {Opt Express}\ }\textbf {\bibinfo {volume} {27}},\ \bibinfo {pages}
    {27369} (\bibinfo {year} {2019})}\BibitemShut {NoStop}%
  \bibitem [{\citenamefont {Ponosova}\ \emph {et~al.}(2022)\citenamefont
    {Ponosova}, \citenamefont {Ruzhitskaya}, \citenamefont {Chaiwongkhot},
    \citenamefont {Egorov}, \citenamefont {Makarov},\ and\ \citenamefont
    {Huang}}]{Ponosova2022}%
    \BibitemOpen
    \bibfield  {author} {\bibinfo {author} {\bibfnamefont {A.}~\bibnamefont
    {Ponosova}}, \bibinfo {author} {\bibfnamefont {D.}~\bibnamefont
    {Ruzhitskaya}}, \bibinfo {author} {\bibfnamefont {P.}~\bibnamefont
    {Chaiwongkhot}}, \bibinfo {author} {\bibfnamefont {V.}~\bibnamefont
    {Egorov}}, \bibinfo {author} {\bibfnamefont {V.}~\bibnamefont {Makarov}},\
    and\ \bibinfo {author} {\bibfnamefont {A.}~\bibnamefont {Huang}},\ }\bibfield
     {title} {\bibinfo {title} {Protecting fiber-optic quantum key distribution
    sources against light-injection attacks},\ }\href
    {https://doi.org/10.1103/PRXQuantum.3.040307} {\bibfield  {journal} {\bibinfo
     {journal} {PRX Quantum}\ }\textbf {\bibinfo {volume} {3}},\ \bibinfo {pages}
    {040307} (\bibinfo {year} {2022})}\BibitemShut {NoStop}%
  \bibitem [{\citenamefont {Lucamarini}\ \emph {et~al.}(2015)\citenamefont
    {Lucamarini}, \citenamefont {Choi}, \citenamefont {Ward}, \citenamefont
    {Dynes}, \citenamefont {Yuan},\ and\ \citenamefont
    {Shields}}]{lucamarini2015}%
    \BibitemOpen
    \bibfield  {author} {\bibinfo {author} {\bibfnamefont {M.}~\bibnamefont
    {Lucamarini}}, \bibinfo {author} {\bibfnamefont {I.}~\bibnamefont {Choi}},
    \bibinfo {author} {\bibfnamefont {M.~B.}\ \bibnamefont {Ward}}, \bibinfo
    {author} {\bibfnamefont {J.~F.}\ \bibnamefont {Dynes}}, \bibinfo {author}
    {\bibfnamefont {Z.~L.}\ \bibnamefont {Yuan}},\ and\ \bibinfo {author}
    {\bibfnamefont {A.~J.}\ \bibnamefont {Shields}},\ }\bibfield  {title}
    {\bibinfo {title} {Practical security bounds against the {T}rojan-horse
    attack in quantum key distribution},\ }\href
    {https://doi.org/10.1103/PhysRevX.5.031030} {\bibfield  {journal} {\bibinfo
    {journal} {Phys. Rev. X}\ }\textbf {\bibinfo {volume} {5}},\ \bibinfo {pages}
    {031030} (\bibinfo {year} {2015})}\BibitemShut {NoStop}%
  \bibitem [{\citenamefont {Gisin}\ \emph {et~al.}(2006)\citenamefont {Gisin},
    \citenamefont {Fasel}, \citenamefont {Kraus}, \citenamefont {Zbinden},\ and\
    \citenamefont {Ribordy}}]{gisin2006}%
    \BibitemOpen
    \bibfield  {author} {\bibinfo {author} {\bibfnamefont {N.}~\bibnamefont
    {Gisin}}, \bibinfo {author} {\bibfnamefont {S.}~\bibnamefont {Fasel}},
    \bibinfo {author} {\bibfnamefont {B.}~\bibnamefont {Kraus}}, \bibinfo
    {author} {\bibfnamefont {H.}~\bibnamefont {Zbinden}},\ and\ \bibinfo {author}
    {\bibfnamefont {G.}~\bibnamefont {Ribordy}},\ }\bibfield  {title} {\bibinfo
    {title} {Trojan-horse attacks on quantum-key-distribution systems},\ }\href
    {https://doi.org/10.1103/PhysRevA.73.022320} {\bibfield  {journal} {\bibinfo
    {journal} {Phys. Rev. A}\ }\textbf {\bibinfo {volume} {73}},\ \bibinfo
    {pages} {022320} (\bibinfo {year} {2006})}\BibitemShut {NoStop}%
  \bibitem [{\citenamefont {Jain}\ \emph {et~al.}(2014)\citenamefont {Jain},
    \citenamefont {Anisimova}, \citenamefont {Khan}, \citenamefont {Makarov},
    \citenamefont {Marquardt},\ and\ \citenamefont {Leuchs}}]{jain2014}%
    \BibitemOpen
    \bibfield  {author} {\bibinfo {author} {\bibfnamefont {N.}~\bibnamefont
    {Jain}}, \bibinfo {author} {\bibfnamefont {E.}~\bibnamefont {Anisimova}},
    \bibinfo {author} {\bibfnamefont {I.}~\bibnamefont {Khan}}, \bibinfo {author}
    {\bibfnamefont {V.}~\bibnamefont {Makarov}}, \bibinfo {author} {\bibfnamefont
    {C.}~\bibnamefont {Marquardt}},\ and\ \bibinfo {author} {\bibfnamefont
    {G.}~\bibnamefont {Leuchs}},\ }\bibfield  {title} {\bibinfo {title}
    {Trojan-horse attacks threaten the security of practical quantum
    cryptography},\ }\href {https://doi.org/10.1088/1367-2630/16/12/123030}
    {\bibfield  {journal} {\bibinfo  {journal} {New J. Phys.}\ }\textbf {\bibinfo
    {volume} {16}},\ \bibinfo {pages} {123030} (\bibinfo {year}
    {2014})}\BibitemShut {NoStop}%
  \bibitem [{\citenamefont {Wu}\ \emph {et~al.}(2020)\citenamefont {Wu},
    \citenamefont {Huang}, \citenamefont {Chen}, \citenamefont {Sun},
    \citenamefont {Ding}, \citenamefont {Qiang}, \citenamefont {Fu},
    \citenamefont {Xu},\ and\ \citenamefont {Wu}}]{wu2020}%
    \BibitemOpen
    \bibfield  {author} {\bibinfo {author} {\bibfnamefont {Z.}~\bibnamefont
    {Wu}}, \bibinfo {author} {\bibfnamefont {A.}~\bibnamefont {Huang}}, \bibinfo
    {author} {\bibfnamefont {H.}~\bibnamefont {Chen}}, \bibinfo {author}
    {\bibfnamefont {S.}~\bibnamefont {Sun}}, \bibinfo {author} {\bibfnamefont
    {J.}~\bibnamefont {Ding}}, \bibinfo {author} {\bibfnamefont {X.}~\bibnamefont
    {Qiang}}, \bibinfo {author} {\bibfnamefont {X.}~\bibnamefont {Fu}}, \bibinfo
    {author} {\bibfnamefont {P.}~\bibnamefont {Xu}},\ and\ \bibinfo {author}
    {\bibfnamefont {J.}~\bibnamefont {Wu}},\ }\bibfield  {title} {\bibinfo
    {title} {Hacking single-photon avalanche detectors in quantum key
    distribution via pulse illumination},\ }\href
    {https://doi.org/10.1364/oe.397962} {\bibfield  {journal} {\bibinfo
    {journal} {Opt Express}\ }\textbf {\bibinfo {volume} {28}},\ \bibinfo {pages}
    {25574} (\bibinfo {year} {2020})}\BibitemShut {NoStop}%
  \bibitem [{\citenamefont {Lydersen}\ and\ \citenamefont
    {Skaar}(2010)}]{lydersen2010}%
    \BibitemOpen
    \bibfield  {author} {\bibinfo {author} {\bibfnamefont {L.}~\bibnamefont
    {Lydersen}}\ and\ \bibinfo {author} {\bibfnamefont {J.}~\bibnamefont
    {Skaar}},\ }\bibfield  {title} {\bibinfo {title} {Security of quantum key
    distribution with bit and basis dependent detector flaws},\ }\href@noop {}
    {\bibfield  {journal} {\bibinfo  {journal} {Quantum Inf. Comput.}\ }\textbf
    {\bibinfo {volume} {10}},\ \bibinfo {pages} {60} (\bibinfo {year}
    {2010})}\BibitemShut {NoStop}%
  \bibitem [{\citenamefont {Lo}\ \emph {et~al.}(2012)\citenamefont {Lo},
    \citenamefont {Curty},\ and\ \citenamefont {Qi}}]{lo2012}%
    \BibitemOpen
    \bibfield  {author} {\bibinfo {author} {\bibfnamefont {H.-K.}\ \bibnamefont
    {Lo}}, \bibinfo {author} {\bibfnamefont {M.}~\bibnamefont {Curty}},\ and\
    \bibinfo {author} {\bibfnamefont {B.}~\bibnamefont {Qi}},\ }\bibfield
    {title} {\bibinfo {title} {Measurement-device-independent quantum key
    distribution},\ }\href {https://doi.org/10.1103/PhysRevLett.108.130503}
    {\bibfield  {journal} {\bibinfo  {journal} {Phys. Rev. Lett.}\ }\textbf
    {\bibinfo {volume} {108}},\ \bibinfo {pages} {130503} (\bibinfo {year}
    {2012})}\BibitemShut {NoStop}%
  \bibitem [{\citenamefont {Siegman}(1962)}]{siegman1962}%
    \BibitemOpen
    \bibfield  {author} {\bibinfo {author} {\bibfnamefont {A.~E.}\ \bibnamefont
    {Siegman}},\ }\bibfield  {title} {\bibinfo {title} {Nonlinear optical
    effects: an optical power limiter},\ }\href
    {https://doi.org/10.1364/AO.1.S1.000127} {\bibfield  {journal} {\bibinfo
    {journal} {Appl Optics}\ }\textbf {\bibinfo {volume} {1}},\ \bibinfo {pages}
    {127} (\bibinfo {year} {1962})}\BibitemShut {NoStop}%
  \bibitem [{\citenamefont {Zhang}\ \emph {et~al.}(2021)\citenamefont {Zhang},
    \citenamefont {Primaatmaja}, \citenamefont {Haw}, \citenamefont {Gong},
    \citenamefont {Wang},\ and\ \citenamefont {Lim}}]{zhang2021}%
    \BibitemOpen
    \bibfield  {author} {\bibinfo {author} {\bibfnamefont {G.}~\bibnamefont
    {Zhang}}, \bibinfo {author} {\bibfnamefont {I.~W.}\ \bibnamefont
    {Primaatmaja}}, \bibinfo {author} {\bibfnamefont {J.~Y.}\ \bibnamefont
    {Haw}}, \bibinfo {author} {\bibfnamefont {X.}~\bibnamefont {Gong}}, \bibinfo
    {author} {\bibfnamefont {C.}~\bibnamefont {Wang}},\ and\ \bibinfo {author}
    {\bibfnamefont {CharlesCiWen.}~\bibnamefont {Lim}},\ }\bibfield  {title} {\bibinfo
    {title} {Securing practical quantum communication systems with optical power
    limiters},\ }\href {https://doi.org/10.1103/PRXQuantum.2.030304} {\bibfield
    {journal} {\bibinfo  {journal} {PRX Quantum}\ }\textbf {\bibinfo {volume}
    {2}},\ \bibinfo {pages} {030304} (\bibinfo {year} {2021})}\BibitemShut
    {NoStop}%
  \bibitem [{\citenamefont {Smith}(1977)}]{smith1977}%
    \BibitemOpen
    \bibfield  {author} {\bibinfo {author} {\bibfnamefont {D.~C.}\ \bibnamefont
    {Smith}},\ }\bibfield  {title} {\bibinfo {title} {High-power laser
    propagation: thermal blooming},\ }\href
    {https://doi.org/10.1109/proc.1977.10809} {\bibfield  {journal} {\bibinfo
    {journal} {Proc. IEEE}\ }\textbf {\bibinfo {volume} {65}},\ \bibinfo {pages}
    {1679} (\bibinfo {year} {1977})}\BibitemShut {NoStop}%
  \bibitem [{\citenamefont {Leite}\ \emph {et~al.}(1967)\citenamefont {Leite},
    \citenamefont {Porto},\ and\ \citenamefont {Damen}}]{leite1967}%
    \BibitemOpen
    \bibfield  {author} {\bibinfo {author} {\bibfnamefont {R.~C.~C.}\
    \bibnamefont {Leite}}, \bibinfo {author} {\bibfnamefont {S.}~\bibnamefont
    {Porto}},\ and\ \bibinfo {author} {\bibfnamefont {T.~C.}\ \bibnamefont
    {Damen}},\ }\bibfield  {title} {\bibinfo {title} {The thermal lens effect as
    a power-limiting device},\ }\href {https://doi.org/10.1063/1.1754849}
    {\bibfield  {journal} {\bibinfo  {journal} {Appl Phys Lett}\ }\textbf
    {\bibinfo {volume} {10}},\ \bibinfo {pages} {100} (\bibinfo {year}
    {1967})}\BibitemShut {NoStop}%
  \bibitem [{\citenamefont {DeRosa}\ and\ \citenamefont
    {Logunov}(2003)}]{derosa2003}%
    \BibitemOpen
    \bibfield  {author} {\bibinfo {author} {\bibfnamefont {M.~E.}\ \bibnamefont
    {DeRosa}}\ and\ \bibinfo {author} {\bibfnamefont {S.~L.}\ \bibnamefont
    {Logunov}},\ }\bibfield  {title} {\bibinfo {title} {Fiber-optic power limiter
    based on photothermal defocusing in an optical polymer},\ }\href
    {https://doi.org/10.1364/ao.42.002683} {\bibfield  {journal} {\bibinfo
    {journal} {Applied optics}\ }\textbf {\bibinfo {volume} {42}},\ \bibinfo
    {pages} {2683} (\bibinfo {year} {2003})}\BibitemShut {NoStop}%
  \bibitem [{\citenamefont {Gottesman}\ \emph {et~al.}(2004)\citenamefont
    {Gottesman}, \citenamefont {Lo}, \citenamefont {L{\" u}tkenhaus},\ and\
    \citenamefont {Preskill}}]{gottesman2004}%
    \BibitemOpen
    \bibfield  {author} {\bibinfo {author} {\bibfnamefont {D.}~\bibnamefont
    {Gottesman}}, \bibinfo {author} {\bibfnamefont {H.-K.}\ \bibnamefont {Lo}},
    \bibinfo {author} {\bibfnamefont {N.}~\bibnamefont {L{\" u}tkenhaus}},\ and\
    \bibinfo {author} {\bibfnamefont {J.}~\bibnamefont {Preskill}},\ }\bibfield
    {title} {\bibinfo {title} {Security of quantum key distribution with
    imperfect devices},\ }\href@noop {} {\bibfield  {journal} {\bibinfo
    {journal} {Quantum Inf. Comput.}\ }\textbf {\bibinfo {volume} {4}},\ \bibinfo
    {pages} {325} (\bibinfo {year} {2004})}\BibitemShut {NoStop}%
  \bibitem [{\citenamefont {Lo}\ \emph {et~al.}(2005)\citenamefont {Lo},
    \citenamefont {Ma},\ and\ \citenamefont {Chen}}]{lo2005}%
    \BibitemOpen
    \bibfield  {author} {\bibinfo {author} {\bibfnamefont {H.-K.}\ \bibnamefont
    {Lo}}, \bibinfo {author} {\bibfnamefont {X.}~\bibnamefont {Ma}},\ and\
    \bibinfo {author} {\bibfnamefont {K.}~\bibnamefont {Chen}},\ }\bibfield
    {title} {\bibinfo {title} {Decoy state quantum key distribution},\ }\href
    {https://doi.org/10.1103/PhysRevLett.94.230504} {\bibfield  {journal}
    {\bibinfo  {journal} {Phys. Rev. Lett.}\ }\textbf {\bibinfo {volume} {94}},\
    \bibinfo {pages} {230504} (\bibinfo {year} {2005})}\BibitemShut {NoStop}%
  \bibitem [{\citenamefont {Ma}\ and\ \citenamefont {Razavi}(2012)}]{ma2012}%
    \BibitemOpen
    \bibfield  {author} {\bibinfo {author} {\bibfnamefont {X.}~\bibnamefont
    {Ma}}\ and\ \bibinfo {author} {\bibfnamefont {M.}~\bibnamefont {Razavi}},\
    }\bibfield  {title} {\bibinfo {title} {Alternative schemes for
    measurement-device-independent quantum key distribution},\ }\href
    {https://doi.org/10.1103/physreva.86.062319} {\bibfield  {journal} {\bibinfo
    {journal} {Phys. Rev. A}\ }\textbf {\bibinfo {volume} {86}},\ \bibinfo
    {pages} {062319} (\bibinfo {year} {2012})}\BibitemShut {NoStop}%
  \bibitem [{\citenamefont {Makarov}(2009)}]{makarov2009}%
    \BibitemOpen
    \bibfield  {author} {\bibinfo {author} {\bibfnamefont {V.}~\bibnamefont
    {Makarov}},\ }\bibfield  {title} {\bibinfo {title} {Controlling passively
    quenched single photon detectors by bright light},\ }\href
    {https://doi.org/10.1088/1367-2630/11/6/065003} {\bibfield  {journal}
    {\bibinfo  {journal} {New J. Phys.}\ }\textbf {\bibinfo {volume} {11}},\
    \bibinfo {pages} {065003} (\bibinfo {year} {2009})}\BibitemShut {NoStop}%
  \bibitem [{\citenamefont {Gras}\ \emph {et~al.}(2020)\citenamefont {Gras},
    \citenamefont {Sultana}, \citenamefont {Huang}, \citenamefont {Jennewein},
    \citenamefont {Bussi{\`e}res}, \citenamefont {Makarov},\ and\ \citenamefont
    {Zbinden}}]{gras2020}%
    \BibitemOpen
    \bibfield  {author} {\bibinfo {author} {\bibfnamefont {G.}~\bibnamefont
    {Gras}}, \bibinfo {author} {\bibfnamefont {N.}~\bibnamefont {Sultana}},
    \bibinfo {author} {\bibfnamefont {A.}~\bibnamefont {Huang}}, \bibinfo
    {author} {\bibfnamefont {T.}~\bibnamefont {Jennewein}}, \bibinfo {author}
    {\bibfnamefont {F.}~\bibnamefont {Bussi{\`e}res}}, \bibinfo {author}
    {\bibfnamefont {V.}~\bibnamefont {Makarov}},\ and\ \bibinfo {author}
    {\bibfnamefont {H.}~\bibnamefont {Zbinden}},\ }\bibfield  {title} {\bibinfo
    {title} {Optical control of single-photon negative-feedback avalanche diode
    detector},\ }\href {https://doi.org/10.1063/1.5140824} {\bibfield  {journal}
    {\bibinfo  {journal} {J Appl Phys}\ }\textbf {\bibinfo {volume} {127}},\
    \bibinfo {pages} {094502} (\bibinfo {year} {2020})}\BibitemShut {NoStop}%
  \bibitem [{\citenamefont {Chistiakov}\ \emph {et~al.}(2019)\citenamefont
    {Chistiakov}, \citenamefont {Huang}, \citenamefont {Egorov},\ and\
    \citenamefont {Makarov}}]{chistiakov2019}%
    \BibitemOpen
    \bibfield  {author} {\bibinfo {author} {\bibfnamefont {V.}~\bibnamefont
    {Chistiakov}}, \bibinfo {author} {\bibfnamefont {A.}~\bibnamefont {Huang}},
    \bibinfo {author} {\bibfnamefont {V.}~\bibnamefont {Egorov}},\ and\ \bibinfo
    {author} {\bibfnamefont {V.}~\bibnamefont {Makarov}},\ }\bibfield  {title}
    {\bibinfo {title} {Controlling single-photon detector id210 with bright
    light},\ }\href {https://doi.org/10.1364/OE.27.032253} {\bibfield  {journal}
    {\bibinfo  {journal} {OE}\ }\textbf {\bibinfo {volume} {27}},\ \bibinfo
    {pages} {32253} (\bibinfo {year} {2019})}\BibitemShut {NoStop}%
  \bibitem [{\citenamefont {Sun}\ \emph {et~al.}(2015)\citenamefont {Sun},
    \citenamefont {Xu}, \citenamefont {Jiang}, \citenamefont {Ma}, \citenamefont
    {Lo},\ and\ \citenamefont {Liang}}]{sun2015}%
    \BibitemOpen
    \bibfield  {author} {\bibinfo {author} {\bibfnamefont {S.-H.}\ \bibnamefont
    {Sun}}, \bibinfo {author} {\bibfnamefont {F.}~\bibnamefont {Xu}}, \bibinfo
    {author} {\bibfnamefont {M.-S.}\ \bibnamefont {Jiang}}, \bibinfo {author}
    {\bibfnamefont {X.-C.}\ \bibnamefont {Ma}}, \bibinfo {author} {\bibfnamefont
    {H.-K.}\ \bibnamefont {Lo}},\ and\ \bibinfo {author} {\bibfnamefont {L.-M.}\
    \bibnamefont {Liang}},\ }\bibfield  {title} {\bibinfo {title} {Effect of
    source tampering in the security of quantum cryptography},\ }\href
    {https://doi.org/10.1103/PhysRevA.92.022304} {\bibfield  {journal} {\bibinfo
    {journal} {Phys. Rev. A}\ }\textbf {\bibinfo {volume} {92}},\ \bibinfo
    {pages} {022304} (\bibinfo {year} {2015})}\BibitemShut {NoStop}%
  \end{thebibliography}%

%

\end{document}